%% file: mainfile.tex
\documentclass[sigconf]{acmart}

\newcommand{\myparatight}[1]{\smallskip\noindent{\bf {#1}:}~}

\usepackage{amsthm}
\usepackage{amsmath}
\usepackage{amssymb}

\usepackage{adjustbox}
\usepackage{booktabs}
\usepackage{array} 
\usepackage{multirow}
\usepackage{graphics}
\usepackage{graphicx}
\usepackage{algorithm}
\usepackage{algorithmic}

\usepackage{amsfonts}
\usepackage{color, url}

\usepackage[skip=3pt]{caption}
\usepackage[skip=3pt]{subcaption}

\usepackage{tabto}
\usepackage{xcolor, colortbl}
\usepackage{enumitem}
\usepackage{bbm}
\usepackage{dsfont}
\usepackage{tcolorbox}

\usepackage{bm}
\usepackage[mathscr]{eucal}
\usepackage[normalem]{ulem}
\useunder{\uline}{\ul}{}
\usepackage{bigstrut,multirow,rotating}
\usepackage{placeins}
\usepackage{pgfplots}
\pgfplotsset{compat=1.17}
\makeatletter
\@namedef{r@tocindent4}{0pt}
\makeatother

\usepackage{xspace}
\newcommand{\alg}{\textsf{RAGForensics}\xspace}

\copyrightyear{2025}
\acmYear{2025}
\setcopyright{acmlicensed}\acmConference[WWW '25]{Proceedings of the ACM Web Conference 2025}{April 28-May 2, 2025}{Sydney, NSW, Australia}
\acmBooktitle{Proceedings of the ACM Web Conference 2025 (WWW '25), April 28-May 2, 2025, Sydney, NSW, Australia}
\acmDOI{10.1145/3696410.3714756}
\acmISBN{979-8-4007-1274-6/25/04}

\begin{document}

\title{Traceback of Poisoning Attacks to Retrieval-Augmented Generation}

\author{Baolei Zhang}
\authornote{Equal contribution.}
\affiliation{%
  \institution{CCS\&CS, DISSec, Nankai University}
  \city{Tianjin}
  \country{China}
  }
\email{zhangbaolei@mail.nankai.edu.cn}

\author{Haoran Xin}
\authornotemark[1]
\affiliation{%
  \institution{CCS\&CS, DISSec, Nankai University}
  \city{Tianjin}
  \country{China}
  }
\email{haoranxin@mail.nankai.edu.cn}

\author{Minghong Fang}
\authornote{Corresponding author.}
\affiliation{%
 \institution{University of Louisville}
\city{Louisville}
 \country{USA}
}
\email{minghong.fang@louisville.edu}

\author{Zhuqing Liu}
\affiliation{%
\institution{University of North Texas}
\city{Denton}
  \country{USA}
}
\email{zhuqing.liu@unt.edu}

\author{Biao Yi}
\affiliation{%
  \institution{CCS\&CS, DISSec, Nankai University}
  \city{Tianjin}
  \country{China}
  }
\email{yibiao@mail.nankai.edu.cn}

\author{Tong Li}
\authornotemark[2]
\affiliation{%
 \institution{CCS\&CS, DISSec, Nankai University}
  \city{Tianjin}
  \country{China}
  }
\email{tongli@nankai.edu.cn}

\author{Zheli Liu}
\affiliation{%
 \institution{CCS\&CS, DISSec, Nankai University}
  \city{Tianjin}
  \country{China}
  }
\email{liuzheli@nankai.edu.cn}

\renewcommand{\shortauthors}{Baolei Zhang et al.}

\begin{abstract}
Large language models (LLMs) integrated with retrieval-augmented generation (RAG) systems improve accuracy by leveraging external knowledge sources. However, recent research has revealed RAG's susceptibility to poisoning attacks, where the attacker injects poisoned texts into the knowledge database, leading to attacker-desired responses. Existing defenses, which predominantly focus on inference-time mitigation, have proven insufficient against sophisticated attacks.  
In this paper, we introduce RAGForensics, the first traceback system for RAG, designed to identify poisoned texts within the knowledge database that are responsible for the attacks. RAGForensics operates iteratively, first retrieving a subset of texts from the database and then utilizing a specially crafted prompt to guide an LLM in detecting potential poisoning texts.  
Empirical evaluations across multiple datasets demonstrate the effectiveness of RAGForensics against state-of-the-art poisoning attacks. This work pioneers the traceback of poisoned texts in RAG systems, providing a practical and promising defense mechanism to enhance their security. Our code is available at: \url{https://github.com/zhangbl6618/RAG-Responsibility-Attribution}

\end{abstract}



\maketitle

\input{introduction}

\input{related}
\input{threatModel}
\input{method}

\input{experiments}

\input{discussion}
\input{conclusion}

\balance
\bibliographystyle{ACM-Reference-Format}
\bibliography{refs}

\input{appendix}

\end{document}

%% file: introduction.tex

\section{Introduction} \label{sec:intro}

Large language models (LLMs)~\cite{brown2020language, achiam2023gpt, anil2023palm} have demonstrated impressive capabilities, matching human-level performance in tasks like question answering and summarization. However, they are prone to hallucinations~\cite{ji2023survey}, generating incorrect information due to the absence of real-time knowledge. This limitation undermines their reliability, particularly in applications where factual accuracy is critical, such as legal, medical, or scientific domains. Retrieval-augmented generation (RAG)~\cite{thoppilan2022lamda, karpukhin2020dense, lewis2020retrieval, borgeaud2022improving, jiang2023active, salemi2024evaluating, chen2024benchmarking, gao2023retrieval} addresses this issue by retrieving relevant texts from an external knowledge database. A RAG system consists of three components: the knowledge database, a retriever, and an LLM. When a user submits a query, the retriever selects the top-$K$ relevant texts from the knowledge database, which are then provided as context to the LLM for generating a more accurate response.

Recent studies~\cite{zou2024poisonedrag, shafran2024machine, zhong2023poisoning, chaudhari2024phantom, deng2024pandora, tan2024glue} highlight the vulnerability of RAG systems to poisoning attacks, where adversaries inject carefully crafted poisoned texts into the knowledge database to manipulate the system’s responses to specific queries. 
For instance, PoisonedRAG~\cite{zou2024poisonedrag} leverages an LLM to craft poisoned texts and injects them into the knowledge database, effectively steering the system toward attacker-specified outputs. 
In response, several defenses~\cite{RobustRAG, zou2024poisonedrag} have been proposed to mitigate the impact of such attacks, primarily by reducing the influence of poisoned texts during inference. For example, RobustRAG~\cite{RobustRAG} employs an isolate-then-aggregate strategy to filter out poisoned keywords from the top-$K$ retrieved texts, thereby enhancing the system’s resilience.

Although existing defenses can mitigate the impact of poisoning attacks on RAG systems to some extent, they remain vulnerable to advanced attacks, where the attacker craft sophisticated strategies to bypass current safeguards~\cite{shan2022poison,deng2024pandora,perez2022ignore,greshake2023not,fang2020local,cao2020fltrust,koh2022stronger,pang2024reconstruction,zhang2024poisoning}. To break this ongoing arms race between attackers and defenders, we focus on identifying the root causes of these attacks and tracing the poisoned texts responsible for them. 
A key step in this direction is integrating poison forensics into RAG systems, which allows the service provider to systematically identify and analyze poisoned texts within the knowledge database. By uncovering compromised data sources and vulnerabilities in the data collection pipeline, poison forensics enables proactive mitigation strategies, such as removing poisoned texts or replacing unreliable sources. This, in turn, enhances the resilience of RAG systems against evolving threats.

However, tracing poisoned texts in RAG systems poses significant challenges. Existing poison forensics techniques in deep learning~\cite{jia2024tracing,rose2024utrace}, which identify poisoned training data by analyzing model gradients or parameters, are inherently unsuitable for RAG systems. Unlike traditional deep learning models, where the service provider has direct access to model parameters, RAG systems rely on third-party-hosted LLMs and retrievers, preventing the direct application of these forensic methods and necessitating alternative tracing approaches.  
Additionally, the vast scale of RAG knowledge databases, often containing millions or billions of entries, makes traceback difficult without inducing high false positive rates. Excessive false positives risk removing benign data, compromising retrieval quality and reliability. Thus, an effective poison forensics solution must accurately detect poisoned texts while minimizing disruptions to legitimate data sources.

In this paper, we conduct the first comprehensive study on tracing poisoned texts in RAG systems under poisoning attacks and introduce \alg, an innovative traceback system for accurately identifying the poisoned texts that are responsible for the attacks.
Our \alg operates iteratively, integrating efficient retrieval with precise identification. Given a user query and its incorrect output reported via feedback, \alg first employs the RAG retriever to retrieve the top-$K$ relevant texts from the knowledge database as potential poisoned candidates. Then, \alg utilizes a specialized prompt to guide an LLM (which may be different from the one used in the RAG system) in determining whether each candidate text is genuinely poisoned. The identified poisoned texts are subsequently removed, updating the knowledge database and serving as the basis for the next iteration. This iterative process continues until the number of identified benign texts reaches $K$. As a result, the final top-$K$ retrieved texts for the user query are exclusively benign, ensuring that the RAG system no longer produces incorrect outputs.

Furthermore, we acknowledge that erroneous responses to user queries, as reported through user feedback, do not necessarily indicate the presence of poisoning attacks. Such inaccuracies, which we refer to as non-poisoned feedback, may instead stem from the LLM acquiring incorrect or imprecise information during its training process. To mitigate this issue, we demonstrate that \alg is capable of effectively distinguishing non-poisoned feedback from poisoned instances. Building on this capability, we introduce a novel benign text enhancement strategy designed to refine and improve the RAG system’s output when faced with non-poisoned feedback, ensuring more accurate and reliable responses.

We summarize our main contributions in this paper as follows:

\begin{list}{\labelitemi}{\leftmargin=1.2em \itemindent=-0.08em \itemsep=.2em}


    \item We introduce \alg, the first traceback framework for the RAG systems, capable of accurately tracing poisoned texts within the knowledge database.

    \item We empirically demonstrate the effectiveness of \alg against three poisoning attacks on three datasets.
    
    \item We design two adaptive attacks against \alg and demonstrate that \alg remains robust against these powerful attacks.

\end{list}

To the best of our knowledge, this study is the first to investigate the traceback of poisoned texts in RAG systems. Our findings highlight the effectiveness of the proposed traceback framework, reinforcing our belief that poison forensics in RAG is both practical and highly promising.

%% file: related.tex

\section{Preliminaries and Related Work} 
\label{sec:related}

\subsection{RAG Overview}

The RAG system improves LLMs by retrieving relevant information from an external knowledge database~\cite{shafran2024machine,chaudhari2024phantom,zou2024poisonedrag,xue2024badrag}. It fetches relevant texts based on a user query and combines them with the query to provide the LLM with additional context, leading to a more accurate response.
The RAG workflow is structured into two stages: knowledge retrieval and answer generation.

\myparatight{Knowledge retrieval}%
The goal of this stage is to retrieve the most relevant texts from the external knowledge database based
on the user’s query. Typically, this is achieved using a vector-based
retrieval model. For a user query $q$, the query encoder $f_q$ generates an embedding vector $v_{q}$. Each text $d_j$ in the knowledge database $ \mathcal{D} $ is encoded into a vector $v_{d_j}$ by a text encoder $f_d$, forming the set $V_\mathcal{D}$. The relevance between $v_{q}$ and each $v_{d_j} \in V_\mathcal{D}$ is measured by similarity (e.g., dot product or cosine similarity). Based on these scores, the top-$K$ most relevant texts $\widehat{\mathcal{R}}(q, K,\mathcal{D})$ are selected.

\myparatight{Answer generation} At this stage, the query $ q $ is combined with the set of retrieved texts $\widehat{\mathcal{R}}(q, K,\mathcal{D})$ to query the LLM, which generates a response $ O $. Formally, the response can be represented as the following:
\begin{align}
O = \text{LLM}(\widehat{\mathcal{R}}(q, K,\mathcal{D}), q ), 
\end{align}
where $\widehat{\mathcal{R}}(q, K, \mathcal{D})$ represents the set of the top-$K$ most relevant texts retrieved from the knowledge database $\mathcal{D}$ based on query $q$. In this stage,  we utilize a similar system prompt as described in~\cite{zou2024poisonedrag} for RAG, as shown in Appendix~\ref{appendix:system_prompt}.

\subsection{Poisoning Attacks to RAG}

The dependence of RAG systems on external data sources creates opportunities to poisoning attacks~\cite{zou2024poisonedrag,shafran2024machine,zhong2023poisoning,xue2024badrag}. In these attacks, the attacker intentionally injects harmful or misleading information into the knowledge database. Their goal is to influence or manipulate the LLM’s responses to specific queries. 
Poisoning attacks to RAG can be implemented by injecting malicious instructions~\cite{greshake2023not} into the knowledge database, inducing the LLM to produce specific responses for targeted queries. In follow-up research, several methods have been proposed to create poisoned texts. For example, Zhong et al.~\cite{zhong2023poisoning} introduced a method to generate adversarial paragraphs that, 
once inserted into the knowledge database, can cause the retriever to retrieve these adversarial passages for specific queries.
Unlike inserting semantically meaningless malicious text, Zou et al.~\cite{zou2024poisonedrag} proposed the PoisonedRAG method, where the attacker injects carefully crafted, semantically meaningful ``poisoned'' texts designed to influence the LLM to generate responses controlled by the attacker.

\subsection{Defenses against Poisoning Attacks to RAG}

Numerous defenses~\cite{fang2024byzantine,fang2022aflguard,fang2025FoundationFL,jagielski2018manipulating,steinhardt2017certified} have been designed to mitigate the effects of poisoning attacks on traditional machine learning systems. However, these strategies are not directly applicable to RAG systems. To address this gap, recent efforts have introduced defenses specifically tailored for RAG systems, including perplexity-based detection~\cite{Jain2023BaselineDF}, query rewriting via the LLM~\cite{Jain2023BaselineDF, zou2024poisonedrag}, and increasing the number of retrieved texts~\cite{zou2024poisonedrag}.
Perplexity-based detection identifies poisoned texts by assessing text quality, assuming that poisoned texts have higher perplexity scores, indicating lower quality. 
Query rewriting defense modifies user queries to reduce the likelihood of retrieving poisoned texts, aiming to retrieve safer, benign information. 
The defense method of increasing the number of retrieved texts effectively reduces the impact of poisoned texts by retrieving more texts.
Additionally, Xiang et al.~\cite{RobustRAG} introduced the RobustRAG defense, which uses an isolate-then-aggregate strategy to defend against such attacks, but its effectiveness declines as the amount of poisoned texts increases.

While current defenses show promise in detecting poisoned texts or preventing them from influencing the outputs of LLMs, stronger and adaptive attackers can still evade them, as seen in deep learning systems~\cite{wenger2021backdoor,yao2019latent,severi2021explanation,bagdasaryan2020blind, schuster2021you}. Drawing inspiration from poison forensics in deep learning~\cite{shan2022poison,jia2024tracing,rose2024utrace}, which focus on tracing poisoned training data associated with a misclassification event, we argue that it is more beneficial for RAG system service providers to prioritize tracing poisoned texts over developing more advanced defenses. However, applying poison forensics to RAG systems presents challenges, as existing methods in deep learning rely on model gradients or parameters, which are generally inaccessible due to limited access to the retriever and LLM.

%% file: threatModel.tex

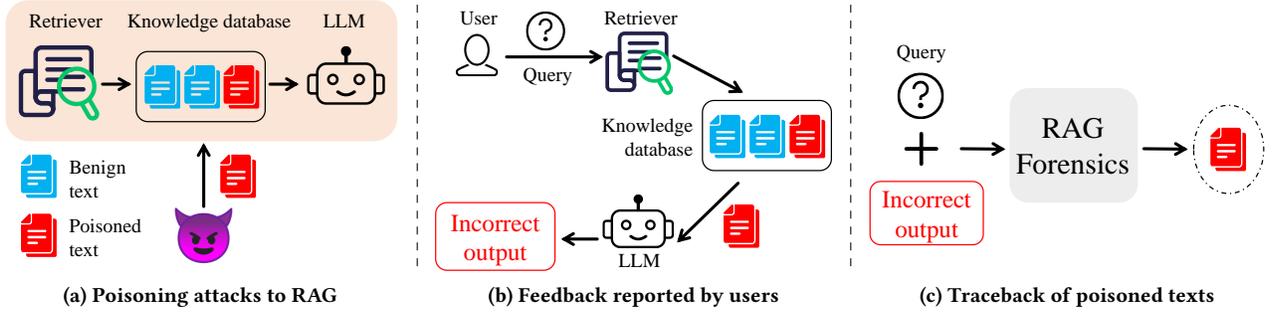
\begin{figure*}[htp]
    \centering
    \begin{tikzpicture}
        \matrix (m) [column sep=0.1cm]
        {
            \node (sub1) {
                \begin{adjustbox}{valign=c}
                    \begin{subfigure}[b]{0.3\textwidth}
                        \centering
                        \includegraphics[width=\textwidth]{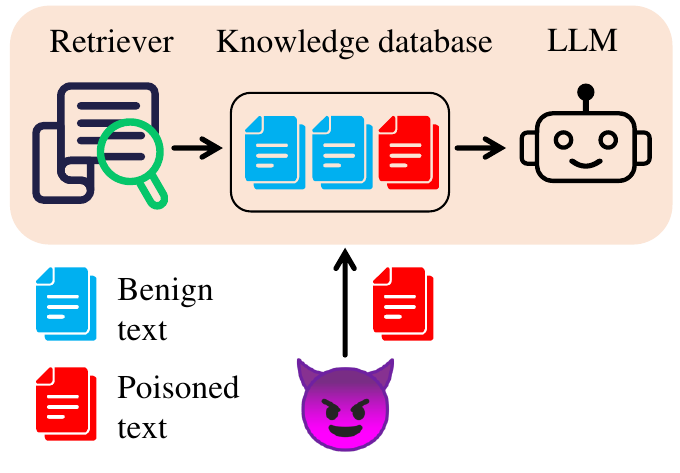} 
                        \caption{Poisoning attacks to RAG}
                        \label{fig:overview_sub1}
                    \end{subfigure}
                \end{adjustbox}
            }; 
            &
            \draw[dashed] (0,-1.5) -- (0,2); 
            &
            \node (sub2) {
                \begin{adjustbox}{valign=c}
                    \begin{subfigure}[b]{0.3\textwidth}
                        \centering
                        \includegraphics[width=\textwidth]{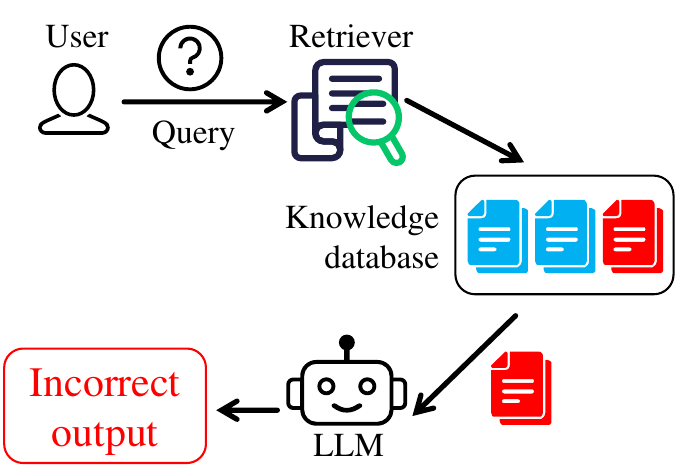} 
                        \caption{Feedback reported by users}
                        \label{fig:overview_sub2}
                    \end{subfigure}
                \end{adjustbox}
            }; 
            &
            \draw[dashed] (0,-1.5) -- (0,2); 
            &
            \node (sub3) {
                \begin{adjustbox}{valign=c}
                    \begin{subfigure}[b]{0.3\textwidth}
                        \centering
                        \includegraphics[width=\textwidth]{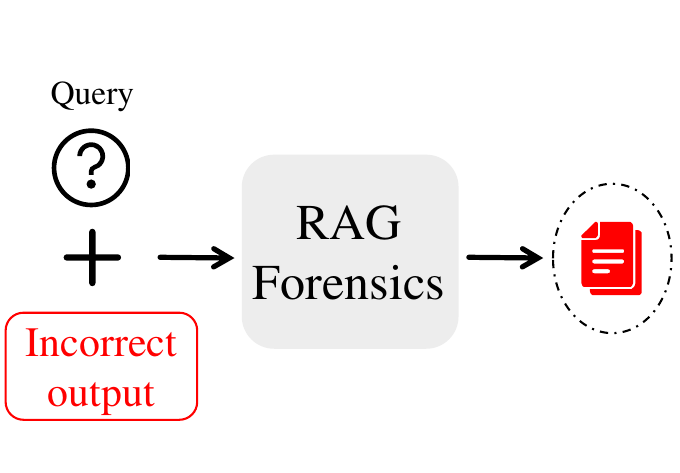} 
                        \caption{Traceback of poisoned texts}
                        \label{fig:overview_sub3}
                    \end{subfigure}
                \end{adjustbox}
            }; 
            \\
        };
    \end{tikzpicture}
    \caption{The example scenario of our traceback system. (a) an attacker injects  poisoned texts into the knowledge database; (b) a user reports feedback that includes the query and the incorrect output; (c) our traceback system \alg identifies poisoned texts based on the user's feedback. }
    \label{fig:overview}
        \vspace{-.15in}
\end{figure*}

\section{Traceback of Poisoned Texts in Poisoning Attacks to RAG}
This paper is the first to address the task of tracing poisoned texts in RAG poisoning attacks. We illustrate this task with the following example scenario.
Figure~\ref{fig:overview} outlines the process of tracing poisoned texts in RAG poisoning attacks. 
The attacker begins by injecting poisoned texts into the knowledge database for a targeted query (Figure~\ref{fig:overview_sub1}). When a user submits a query similar to the targeted query, the RAG system, relying on the poisoned database, generates an incorrect output aligned with the attacker’s intent (Figure~\ref{fig:overview_sub2}).
To address this, the RAG service provider can offer a feedback button, allowing users to report the incorrect output along with the user query. This report is sent to the traceback system, which identifies the poisoned texts in the knowledge database responsible for the incorrect output (Figure~\ref{fig:overview_sub3}).
In what follows,
we begin by outlining the threat model for poisoning attacks and the traceback system in RAG. Afterward, we analyze the key challenges involved in designing an effective traceback system.

\subsection{Threat Model} 
\label{sec:threat_model}

In this section, we describe the threat model, along with the assumptions underlying both the attacker and the traceback system.

\myparatight{Attacker}%
Building on existing poisoning attacks~\cite{zou2024poisonedrag, shafran2024machine, zhong2023poisoning, xue2024badrag}, we outline the attacker’s goal and knowledge. 
The attacker aims to poison the knowledge database so that the LLM in RAG generates a specific, attacker-chosen response to a targeted query.
We assume the attacker has full knowledge of the texts in the database and direct access to the parameters of both the retriever and the LLM, allowing them to query these components directly.

\myparatight{Traceback system}%
We assume that the service provider of the traceback system is the owner of the RAG system. The traceback system is granted full access to all texts in the knowledge database. However, for the retriever and LLM, we consider a practical scenario where the RAG owner uses a closed-source retriever and LLM. As a result, the traceback system cannot access their internal parameters but can query them directly.
We assume that the traceback system has collected a set of user queries and their incorrect RAG outputs as reported by users.
This is a common assumption in traceback systems for poisoning attacks~\cite{shan2022poison, jia2024tracing}. Many LLM applications, such as ChatGPT\footnote{https://chatgpt.com/}, include a feedback button that allows users to report incorrect outputs along with the related queries. Thus, this assumption is both practical and easily implementable in RAG.

Note that in practice, incorrect outputs can stem from the LLM itself, as under-training or incorrect knowledge can lead to inaccuracies even without poisoned texts.
In Section~\ref{sec:discussion}, we explore how to identify whether incorrect outputs reported via user feedback are caused by poisoning attacks, and propose a post-hoc defense to improve its accuracy across queries.

\subsection{Design Challenges}
\label{sec:challenge}

\myparatight{Optimization problem}
Building on the traceback of data poisoning attacks in neural networks~\cite{shan2022poison}, we formulate the traceback of poisoned texts in RAG poisoning attacks as an optimization problem. 
Unlike poisoning attacks in neural networks, where model parameters are manipulated by injecting poisoned data into the training dataset, RAG poisoning attacks involve injecting multiple poisoned texts into the knowledge database to induce the LLM to generate attacker-desired answers in response to targeted queries. The goal of the traceback process is to identify the poisoned texts in the knowledge database responsible for the incorrect outputs. 
Given a set of independent user queries $\mathcal{Q}$ and their corresponding incorrect outputs (collected through user feedback), our goal is to identify a subset of poisoned texts $\mathcal{D}_p$ within the poisoned knowledge database $\mathcal{D}$. The objective is to ensure that after removing $\mathcal{D}_p$ from $\mathcal{D}$, the LLM no longer generates the incorrect output $t_i$ for each user query $q_i \in \mathcal{Q}$. 
The optimization problem can be formalized as follows:
\begin{align}
\label{optimization_problem}
\min_{\mathcal{D}_p} & \quad \frac{1}{|\mathcal{Q}|} \sum_{i=1}^{|\mathcal{Q}|} \mathbb{I} ( \text{LLM}( \widehat{\mathcal{R}}(q_i, K,\mathcal{D}\setminus \mathcal{D}_p), q_i ) \xrightarrow{\text{Match}} t_i ) \\ \nonumber 
\text{s.t.} & \quad q_i \in \mathcal{Q}, \quad i=1,2,\dots,|\mathcal{Q}|,  \\ \nonumber 
& \quad \mathcal{D}_p \in \mathcal{D}, \nonumber
\end{align} 
where $|\mathcal{Q}|$ denotes the number of user queries, $\mathcal{D} \setminus \mathcal{D}_p$ represents the knowledge database after removing the identified poisoned texts $\mathcal{D}_p$ from $\mathcal{D}$, 
$\widehat{\mathcal{R}}(q_i, K, \mathcal{D} \setminus \mathcal{D}_p)$ represents the top-$K$ texts retrieved from the database $\mathcal{D} \setminus \mathcal{D}_p$ for the query $q_i$,
and $\mathbb{I}(\cdot)$ is an indicator function that returns 1 if the LLM's output matches (or aligns with) the incorrect output $t_i$, and 0 otherwise.

\myparatight{Key challenges} %
The main challenge in identifying the poisoned texts $\mathcal{D}_p$ lies in solving the optimization problem in Eq.~(\ref{optimization_problem}). A straightforward approach, inspired by~\cite{shan2022poison}, would be to iteratively update $\mathcal{D}_p$ using the gradient of the objective function. However, this is impractical under our threat model.
First, computing the gradient is challenging due to two key constraints: (1) we lack access to the parameters of both the retriever and the LLM, and (2) both $\widehat{\mathcal{R}}(\cdot)$ and $\mathbb{I}(\cdot)$ are discrete functions, making gradient-based optimization infeasible.
Second, determining update candidates for $\mathcal{D}_p$ would require computing the gradient for every text in the knowledge database $\mathcal{D}$, introducing substantial computational overhead and further complicating the optimization process.

%% file: method.tex

\section{Our Traceback System: \alg}

\subsection{Overview}
\label{subsec_overview}

To address the key challenge in Section \ref{sec:challenge}, our \alg identifies the poisoned texts $\mathcal{D}_p$ by iteratively retrieving and identifying them for each user query. For each query, \alg retrieves texts likely to be poisoned and identifies those responsible for the incorrect output. The process stops once no poisoned texts remain among the top-$K$ relevant texts. 

\begin{algorithm}[!t]
   \caption{\alg.}
   \label{alg:main}
\begin{algorithmic}[1]
   \STATE {\bfseries Input:} The set of user queries $\mathcal{Q}$, incorrect output $t_i$ for query $q_i \in \mathcal{Q}$, poisoned knowledge database $\mathcal{D}$, the value of $K$.
   \STATE {\bfseries Output:} The set of poisoned texts $\mathcal{D}_p$.
    \STATE Initialize $\mathcal{D}_p \gets \emptyset$.
   \FOR{$i = 1$ to $|\mathcal{Q}|$}
   \STATE Initialize $\mathcal{C}_p \gets \emptyset$, $\mathcal{C}_b \gets \emptyset$.
   \WHILE{$|\mathcal{C}_b| < K$}
      \STATE $\mathcal{C} \gets   \widehat{\mathcal{R}}(q_i, K,\mathcal{D}\setminus \mathcal{C}_p)$
      \FOR{$j = 1$ to $|\mathcal{C}|$}
      \STATE $\Gamma(\mathcal{C}_j) =  \text{LLM}(q_i,\mathcal{C}_j,t_i)$      
      \IF{$\Gamma(\mathcal{C}_j)$ contains ``[Label: Yes]''}
      \STATE $\mathcal{C}_p \gets \mathcal{C}_p \cup \{\mathcal{C}_j\}$
      \ELSE
      \STATE $\mathcal{C}_b \gets \mathcal{C}_b \cup \{\mathcal{C}_j\}$
      \ENDIF
      \ENDFOR
   \ENDWHILE
   \STATE $\mathcal{D}_p \gets \mathcal{D}_p \cup \mathcal{C}_p$
   \ENDFOR
   \STATE \textbf{return} $\mathcal{D}_p$
\end{algorithmic}
\end{algorithm}

\subsection{\alg}

\label{sec:alg}

Since the traceback system lacks knowledge of the attacker's strategy for injecting poisoned texts, we assume a worst-case scenario where the poisoned texts are randomly distributed throughout the knowledge database.
Consequently, given a user query \( q_i \in \mathcal{Q} \), a naive approach would be to exhaustively search the poisoned knowledge database \( \mathcal{D} \) to find poisoned texts responsible for the incorrect output \( t_i \). However, this method is computationally prohibitive and prone to false positives, potentially misclassifying benign texts. To mitigate these issues, we introduce an iterative method designed to efficiently locate and identify poisoned texts while reducing the false positive rates.

Our \alg builds on the key observation that the attacker crafts poisoned texts to closely resemble the user query \( q_i \) in the embedding space, ensuring their retrieval among the top-\( K \) relevant texts. The presence of an incorrect output \( t_i \) further confirms that poisoned texts have been retrieved. Based on this insight, our method employs the RAG retriever to iteratively retrieve the top-\( K \) relevant texts for \( q_i \) as potential poisoned candidates.  
Specifically, given a query \( q_i \), let \( \mathcal{C} \) denote the top-\( K \) texts retrieved from the current knowledge database after removing previously identified poisoned texts, where \( |\mathcal{C}| = K \). We denote the \( j \)-th text in \( \mathcal{C} \) as \( \mathcal{C}_j \). The core challenge then lies in determining whether \( \mathcal{C}_j \) is poisoned by assessing whether it triggers the incorrect output \( t_i \).  
However, due to the inherent complexity of natural language, designing a deterministic function for precise identification is difficult. To address this, we leverage an LLM for classification, following a common NLP approach. This method, however, introduces two key challenges. First, while LLMs exhibit strong text comprehension abilities, they require explicit instructions to assess whether a text is poisoned. Second, since the incorrect output \( t_i \) is often factually incorrect or harmful, modern safety-aligned LLMs may activate their safety mechanisms upon encountering poisoned texts, potentially interfering with the classification process.

To overcome these challenges, we design a structured prompt to guide the LLM in determining whether the text \( \mathcal{C}_j \) contributes to the incorrect output \( t_i \). Our prompt explicitly instructs the LLM to disregard both background knowledge and the factual correctness of \( \mathcal{C}_j \) and \( t_i \), ensuring a focus solely on whether the retrieved text influences the incorrect output. Additionally, we integrate a step-by-step reasoning process inspired by the chain-of-thought (CoT) approach \cite{wei2022chain} to enhance the reliability of the LLM’s judgment.
The prompt follows a three-step structure to systematically guide the LLM’s evaluation. First, it directs the LLM to assess whether \( \mathcal{C}_j \) encourages an answer aligned with \( t_i \), irrespective of factual accuracy. Second, it instructs the LLM to provide a brief explanation for its assessment, fostering logical reasoning and improving reliability. Lastly, we incorporate the CoT technique by explicitly including the phrase, ``Let’s think step by step'', prompting the LLM to conduct a structured, systematic analysis. The full prompt is presented below:

\begin{center}

\begin{tcolorbox}[colback=gray!10,
                  colframe=black!80,
                  width=\linewidth,
                  arc=1mm, auto outer arc,
                  boxrule=1pt,
                  title =  The prompt designed to guide the LLM in assessing whether the text \( \mathcal{C}_j \) is poisoned.
                 ]
I will provide a question, an response, and a context. \\
Without any other knowledge, judge whether the provided context tries to induce you to generate an answer consistent with the provided response [$t_i$], regardless of whether it is correct. \\
Begin your judgement by providing a short explanation. \\
If yes, return ``[Label: Yes]'', otherwise, return ``[Label: No]''. \\
Question: [$q_i$] \\
Context: [$\mathcal{C}_j$] \\
Response:[$t_i$] \\
Let's think step by step.

\end{tcolorbox}
\end{center}

Specifically, we substitute the relevant placeholders in the prompt with \( q_i \), \( \mathcal{C}_j \), and \( t_i \) as input to the LLM, obtaining its output \( \Gamma(\mathcal{C}_j) \). If \( \Gamma(\mathcal{C}_j) \) contains the string ``[Label: Yes]'', it indicates that the LLM has identified \( \mathcal{C}_j \) as responsible for the incorrect output \( t_i \), classifying it as a poisoned text, which we then add to the detected poisoned text set \( \mathcal{C}_p \). Conversely, if \( \Gamma(\mathcal{C}_j) \) contains ``[Label: No]'', the LLM determines that \( \mathcal{C}_j \) does not contribute to \( t_i \), and we add it to the benign text set \( \mathcal{C}_b \).
This process iterates until the number of identified benign texts reaches \( K \), ensuring that once all detected poisoned texts are removed from the knowledge database, the top-\( K \) retrieved texts for the user query \( q_i \) consist solely of benign content. As a result, the RAG system is prevented from generating the incorrect output \( t_i \).

Our \alg eliminates the need for exhaustive database searches while enhancing the precision of poisoned text identification through structured LLM evaluation. Even if \( t_i \) contains malicious content that might hinder the LLM's assessment, our carefully designed prompt maintains effectiveness by providing clear instructions and step-by-step guidance.
The pseudocode of our proposed \alg algorithm is presented in Algorithm~\ref{alg:main}.

\begin{table*}[t]
\setlength\tabcolsep{7pt}
\caption{The DACCs, FPRs and FNRs of our \alg and 6 traceback baselines against 3 poisoning attacks on 3 datasets. Bold font indicates the best results, while underlined font represents the second-best results. All values are expressed as percentages.}
\label{tab:traceback_main_result}
\begin{tabular}{c|c|c|ccccccc}
\toprule[1pt]
Datasets & Attacks & Metrics & PPL-100 & PPL-90 & ExpGen & RKM & TKM & PoiFor & \alg \\ \hline
\multirow{9}{*}{NQ} & \multirow{3}{*}{PoisonedRAG-B} & DACC $\uparrow $ & 37.5 & 37.5 & 90.3 & 84.3 & 81.0 & 83.1 & \textbf{99.6} \\
 &  & FPR $\downarrow $ & 0.0 & 0.0 & 0.0 & 7.6 & 34.1 & 2.2 & {\ul 0.8} \\
 &  & FNR $\downarrow $ & 100.0 & 100.0 & 19.4 & 23.9 & 3.9 & 31.5 & \textbf{0.0} \\ \cline{2-10} 
 & \multirow{3}{*}{PoisonedRAG-W} & DACC $\uparrow $ & 16.7 & 16.7 & 91.2 & 84.9 & 72.9 & 81.1 & \textbf{99.2} \\
 &  & FPR $\downarrow $ & 0.0 & 0.0 & 0.5 & 8.8 & 38.8 & 0.0 & 1.6 \\
 &  & FNR $\downarrow $ & 100.0 & 100.0 & 17.1 & 24.1 & 15.4 & 37.9 & \textbf{0.0} \\ \cline{2-10} 
 & \multirow{3}{*}{InstruInject} & DACC $\uparrow $ & 28.6 & 28.6 & 100.0 & 69.6 & 86.3 & 100.0 & {\ul 99.6} \\
 &  & FPR $\downarrow $ & 0.0 & 0.0 & 0.0 & 0.8 & 11.5 & 0.0 & {\ul 0.4} \\
 &  & FNR $\downarrow $ & 100.0 & 100.0 & 0.0 & 60.0 & 16.0 & 0.0 & {\ul 0.4} \\ \hline
\multirow{9}{*}{HotpotQA} & \multirow{3}{*}{PoisonedRAG-B} & DACC $\uparrow $ & 0.0 & 68.2 & 87.5 & 77.7 & 85.4 & 75.5 & \textbf{97.4} \\
 &  & FPR $\downarrow $ & 0.0 & 18.1 & 2.6 & 6.7 & 29.1 & 19.6 & {\ul 2.4} \\
 &  & FNR $\downarrow $ & 100.0 & 45.6 & 22.5 & 37.8 & 0.0 & 29.5 & {\ul 2.8} \\ \cline{2-10} 
 & \multirow{3}{*}{PoisonedRAG-W} & DACC $\uparrow $ & 0.0 & 60.1 & 89.2 & 80.3 & 64.1 & 75.1 & \textbf{97.6} \\
 &  & FPR $\downarrow $ & 85.6 & 79.7 & 1.6 & 7.9 & 35.4 & 21.7 & \textbf{1.6} \\
 &  & FNR $\downarrow $ & 57.2 & 0.0 & 20.0 & 31.4 & 36.4 & 44.4 & {\ul 3.2} \\ \cline{2-10} 
 & \multirow{3}{*}{InstruInject} & DACC $\uparrow $ & 15.6 & 60.5 & 99.1 & 68.6 & 87.5 & 98.9 & 98.2 \\
 &  & FPR $\downarrow $ & 3.3 & 79.1 & 1.8 & 0.7 & 8.9 & 2.2 & 2.3 \\
 &  & FNR $\downarrow $ & 98.0 & 0.0 & 0.0 & 62.1 & 16.0 & 0.0 & {\ul 1.2} \\ \hline
\multirow{9}{*}{MS-MARCO} & \multirow{3}{*}{PoisonedRAG-B} & DACC $\uparrow $ & 44.4 & 67.6 & 83.4 & 76.6 & 74.4 & 73.0 & \textbf{98.4} \\
 &  & FPR $\downarrow $ & 0.0 & 8.3 & 0.0 & 18.1 & 49.2 & 1.0 & 2.3 \\
 &  & FNR $\downarrow $ & 100.0 & 56.4 & 33.3 & 28.7 & 2.0 & 53.0 & \textbf{0.8} \\ \cline{2-10} 
 & \multirow{3}{*}{PoisonedRAG-W} & DACC $\uparrow $ & 69.5 & 64.1 & 87.8 & 78.5 & 40.2 & 66.6 & \textbf{98.3} \\
 &  & FPR $\downarrow $ & 0.0 & 71.7 & 0.0 & 17.9 & 47.7 & 3.5 & {\ul 2.7} \\
 &  & FNR $\downarrow $ & 61.0 & 0.0 & 24.4 & 25.1 & 71.9 & 63.3 & {\ul 0.8} \\ \cline{2-10} 
 & \multirow{3}{*}{InstruInject} & DACC $\uparrow $ & 86.2 & 64.1 & 99.2 & 53.8 & 73.7 & 97.8 & \textbf{99.4} \\
 &  & FPR $\downarrow $ & 0.0 & 71.7 & 0.0 & 8.7 & 11.4 & 2.5 & {\ul 1.2} \\
 &  & FNR $\downarrow $ & 27.6 & 0.0 & 1.6 & 83.8 & 41.2 & 2.0 & \textbf{0.0} \\ 
 \bottomrule[1pt]
\end{tabular}
    \vspace{-.15in}
\end{table*}

%% file: experiments.tex

\section{Experiments} \label{sec:exp}

\subsection{Experimental Setup}

\subsubsection{Datasets} 
We utilize three question-answering datasets: Natural Questions (NQ)~\cite{Kwiatkowski2019NaturalQA}, MS-MARCO~\cite{Campos2016MSMA}, and HotpotQA~\cite{Yang2018HotpotQAAD}. Each dataset contains a collection of queries with an associated knowledge database. For each query, multiple ground truth texts are provided to facilitate answer generation. NQ dataset originates from real Google search queries, with Wikipedia pages serving as its knowledge database. Similarly, MS-MARCO dataset is built upon Bing search queries, using retrieved web pages from Bing as its knowledge source. HotpotQA dataset consists of human-crafted questions that require multi-hop reasoning, and its knowledge database is also derived from Wikipedia.

In our experiments, we collect 50 incorrect events (feedback reported by users) for each attack performed on each dataset. Specifically, we first perform a poisoning attack by injecting poisoned texts into the knowledge database. Then, we use all targeted queries as user queries to obtain outputs of the RAG system. Finally, we collect 50 queries whose RAG outputs match the attacker-desired answers, along with their corresponding RAG outputs, as the final incorrect events.

\subsubsection{Attacks}
To assess the effectiveness of~\alg, we employ the following poisoning attacks:

\myparatight{PoisonedRAG attack~\cite{zou2024poisonedrag}} PoisonedRAG aims to inject $M$ poisoned texts into the knowledge database such that RAG generates the attacker-desired answer for the targeted query. Each poisoned text $P$ is divided into two subtexts, $S$ and $I$, where $P = S \oplus I$, with $\oplus$ denoting string concatenation. For subtext $I$, the attacker uses an LLM to generate it in a way that ensures the LLM produces the attacker-desired answer when $I$ is used as context. For subtext $S$, various techniques are employed depending on whether the retriever operates in a black-box or white-box setting.

\begin{list}{\labelitemi}{\leftmargin=1.2em \itemindent=-0.08em \itemsep=.2em}
\item \myparatight{Black-box (PoisonedRAG-B)} The attacker uses only the targeted query as subtext $S$.
\item \myparatight{White-box (PoisonedRAG-W)} The attacker initially sets the targeted query as subtext $S$ and then updates it to maximize the similarity score between subtext $S$ and the targeted query in the embedding space.
\end{list}

\myparatight{Instruction injection attack (InstruInject)~\cite{shafran2024machine}} 
The poisoned text for each targeted query is crafted in the same manner as in PoisonedRAG (black-box setting). However, the subtext $I$ is replaced with the malicious instruction designed to induce the LLM to generate the attacker-desired answer.

\subsubsection{Traceback Baselines in Poisoning Attacks to RAG}
\label{subsubsec:traceback_baseline}
Given the limited research on traceback methods for poisoning attacks in RAG, there are no established traceback methods that can serve as baselines. 
To provide a more comprehensive evaluation of~\alg, we introduce six baselines by extending traceback methods from poisoning attacks on neural networks and adapting commonly used defenses against poisoning attacks in RAG systems.

\myparatight{Perplexity-based detection (PPL)}
Perplexity-based detection, as proposed in~\cite{zou2024poisonedrag, shafran2024machine}, provides a defense mechanism against poisoning attacks targeting RAG. In our experiment, we extend this approach to serve as a traceback baseline. Specifically, for each dataset, we randomly select 1,000 texts from the knowledge database and calculate the perplexity of each text using Llama-2-7b~\cite{touvron2023llama}. Among the retrieved top-$K$ texts, we identify those with perplexity values exceeding the predetermined threshold as poisoned texts. 
We further establish two baselines based on threshold values:
\begin{list}{\labelitemi}{\leftmargin=1.2em \itemindent=-0.08em \itemsep=.2em}
   \item \myparatight{PPL-100}The threshold is chosen such that it exceeds the perplexity of all texts.
\item \myparatight{PPL-90}The threshold is chosen such that it exceeds the perplexity of 90\% of the texts.
\end{list}

\myparatight{Explanation generation (ExpGen)} 
Prior work~\cite{cuconasu2024power} has suggested that LLM-generated explanations can enhance source transparency. Building on this foundation, we modify the RAG system prompt to instruct the LLM to explain which specific texts from the retrieved top-$K$ texts are used to generate the answer. We then identify poisoned texts by examining which texts are referenced in the LLM's explanations when generating incorrect answers.

\myparatight{Response keywords matching (RKM)}
We adapt RobustRAG~\cite{RobustRAG}, originally designed as a defense mechanism against RAG poisoning attacks, into a traceback method. RKM first prompts the RAG's LLM with each text from the retrieved top-$K$ texts to extract keywords from the responses, following RobustRAG's methodology. RKM then identifies texts as poisoned through substring matching between these extracted keywords and the incorrect output $t_i$.

\myparatight{Text keywords matching (TKM)}
Unlike the RKM baseline, TKM directly extracts the keywords from each text in the retrieved top-$K$ texts.

\myparatight{Poison Forensics (PoiFor)}
Poison Forensics~\cite{shan2022poison} is a traceback mechanism designed to address poisoning attacks on neural networks. It works by iteratively clustering and pruning benign training data to trace a set of poisoned data. 
When adapted for RAG, PoiFor utilizes an LLM to cluster the retrieved top-$K$ texts for each query $q_i$ into two groups. It then prompts the LLM with texts from each cluster to identify the poisoned one—the cluster that produces outputs aligned with the incorrect output $t_i$.

\subsubsection{Evaluation Metric} 
To evaluate our traceback system, we assess identification performance using the following metrics: detection accuracy (DACC), false positive rate (FPR), false negative rate (FNR), attack success rate (ASR), and accuracy rate (ACC). The detailed computations of them are provided in Appendix~\ref{appendix:metrics}.

\begin{figure*}[t]
\centering
\subfloat{\includegraphics[height=4cm]{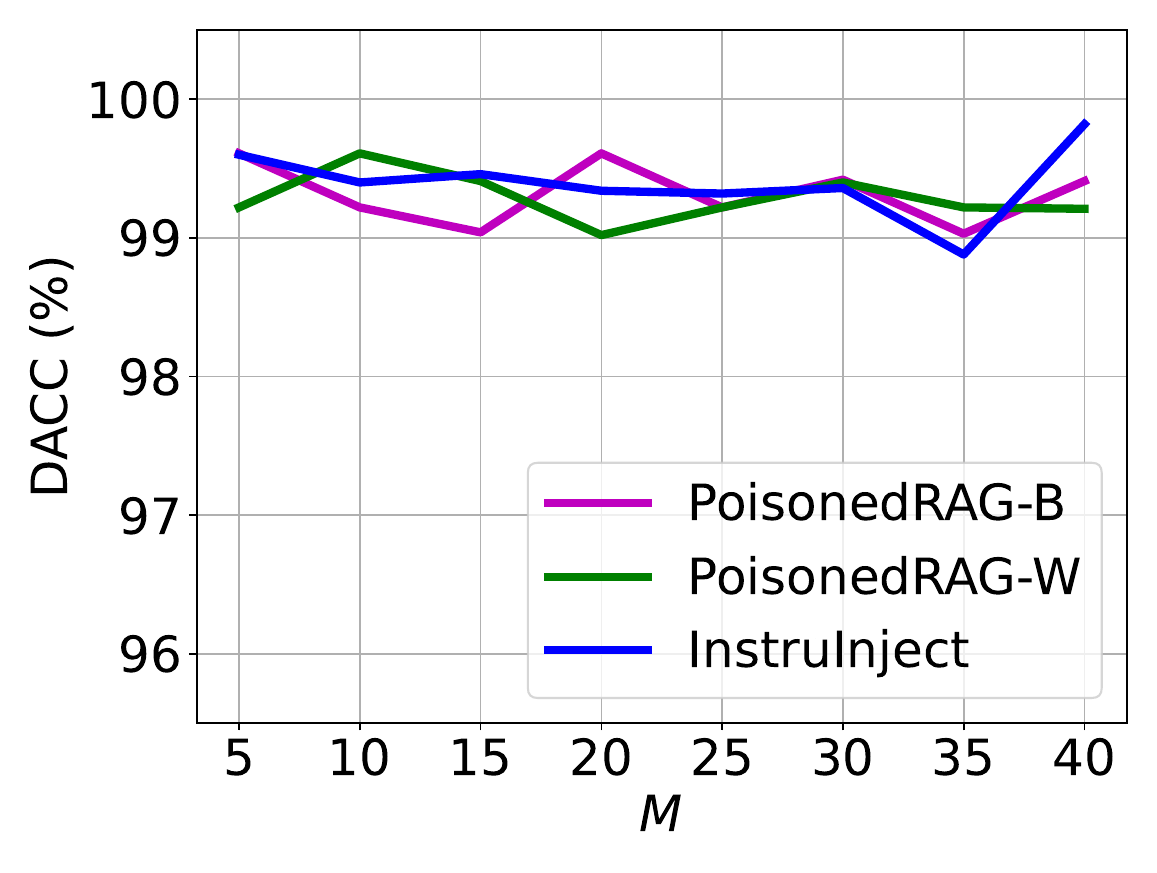}}
\subfloat{\includegraphics[height=4cm]{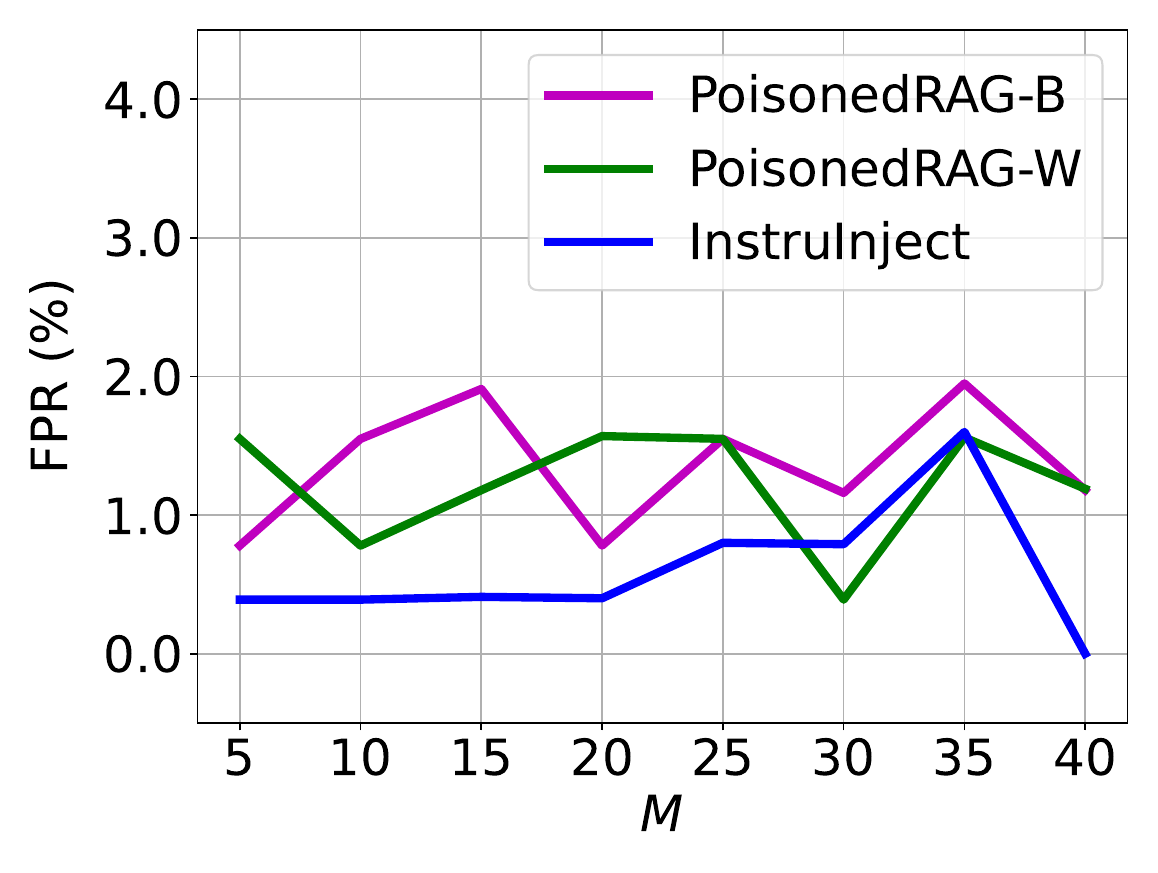}}
\subfloat{\includegraphics[height=4cm]{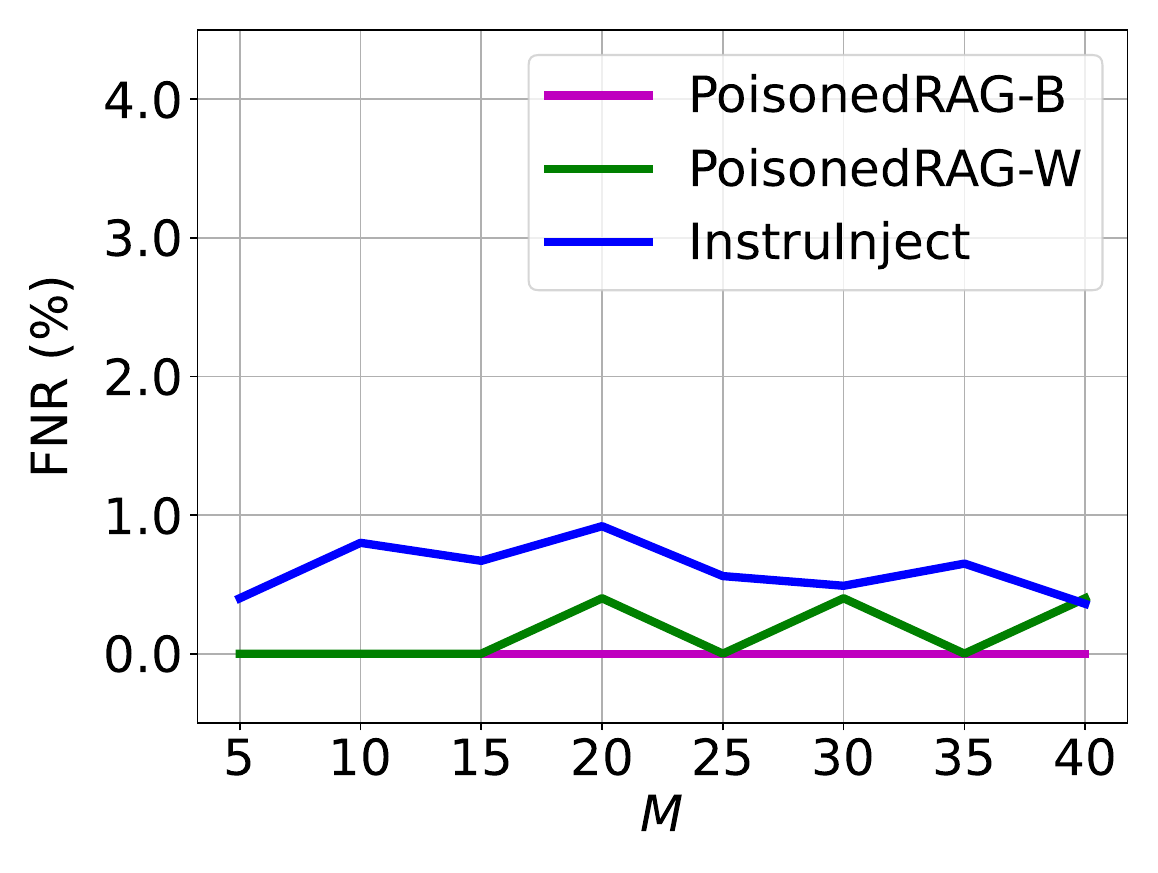}}

\caption{Impact of the number of poisoned texts for each targeted query on NQ dataset.}
\label{fig:traceback_m}
    \vspace{-.15in}
\end{figure*}

\subsubsection{Parameter Setting}
We outline the default configurations for the RAG system, the attack settings, and our traceback system. To reduce computational costs, we conduct 5 iterations for each experiment, with each iteration randomly selecting 10 queries for evaluation.

\myparatight{RAG settings} For the retriever, we use Contriever~\cite{Izacard2021UnsupervisedDI} by default and apply the dot product to calculate the similarity score. We retrieve the $K=5$ most relevant texts from the knowledge database for each query. For the LLM component, we use GPT-4o-mini as the default LLM in RAG.

\myparatight{Poisoning attack settings} In general, we set the number of poisoned texts per targeted query, denoted as $M$, to 5. For the PoisonedRAG attack, we use GPT-4o-mini to craft the poisoned texts.

\myparatight{Our traceback system settings} We employ GPT-4o-mini as the default LLM to determine whether each potential poisoned text candidate is responsible for the incorrect output $t_i$.

\subsection{Experimental Results}
\myparatight{Our \alg can accurately trace the poisoned texts of various poisoning attacks} Table \ref{tab:traceback_main_result} presents the DACC, FPR, and FNR metrics of \alg when tested against three poisoning attacks across three datasets. We observe the following key findings. First, \alg consistently identifies all poisoned texts within the knowledge database with high accuracy: the DACCs exceed 97.4\%, the FPRs remain below 2.7\%, and the FNRs stay below 3.2\%. Second, \alg demonstrates stable performance across different poisoning attacks. For example, in the NQ dataset, the DACCs vary by no more than 0.4\%, the FPRs by no more than 1.2\%, and the FNRs by no more than 0.4\%.

\myparatight{Our \alg outperforms all traceback baselines} Table \ref{tab:traceback_main_result} also presents a comparison between \alg and other baselines across various poisoning attacks. The results demonstrate that \alg significantly outperforms all other traceback baselines. Moreover, \alg exhibits superior generalization performance, while other baselines only achieve good performance in limited settings with specific poisoning attacks and datasets.

\begin{table}[t]
\setlength\tabcolsep{3pt}
\caption{Impact of LLM used to identify the poisoned texts in our \alg on NQ dataset.}
\label{tab:traceback_llm}
\begin{tabular}{c|c|ccc}
\toprule[1pt]
Attacks & Metrics & GPT-4o-mini & GPT-4-turbo & GPT-4o \\ \hline
\multirow{3}{*}{PoisonedRAG-B} & DACC $\uparrow $ & 99.6 & 97.4 & 99.4 \\
 & FPR $\downarrow $ & 0.8 & 4.1 & 0.8 \\
 & FNR $\downarrow $ & 0.0 & 1.2 & 0.4 \\ \hline
\multirow{3}{*}{PoisonedRAG-W} & DACC $\uparrow $ & 99.2 & 99.6 & 99.6 \\
 & FPR $\downarrow $ & 1.6 & 0.4 & 0.0 \\
 & FNR $\downarrow $ & 0.0 & 0.4 & 0.8 \\ \hline
\multirow{3}{*}{InstruInject} & DACC $\uparrow $ & 99.6 & 97.5 & 98.0 \\
 & FPR $\downarrow $ & 0.4 & 5.0 & 0.8 \\
 & FNR $\downarrow $ & 0.4 & 0.0 & 3.2 \\
 \bottomrule[1pt]
\end{tabular}
    \vspace{-.15in}
\end{table}

\myparatight{Impact of the number of poisoned texts per targeted query} 
We evaluate \alg's effectiveness as the number of injected poisoned texts per targeted query (denoted as $M$) varies from 5 to 40. Figure \ref{fig:traceback_m} presents the results on the NQ dataset, while results for HotpotQA and MS-MARCO datasets are shown in Figures~\ref{appendix:traceback_m_hotpotqa} and \ref{appendix:traceback_m_msmarco} in Appendix. The results demonstrate that \alg maintains consistent performance across different numbers of poisoned texts, indicating its robustness to varying attack scales.

\myparatight{Impact of LLMs in identifying poisoned texts in \alg} We conduct experiments using different LLMs in \alg to identify poisoned texts, including GPT-4o, GPT-4-turbo, and GPT-4o-mini. The results on the NQ dataset are presented in Table \ref{tab:traceback_llm}, with results for HotpotQA and MS-MARCO datasets provided in Table~\ref{appendix:traceback_llm} in Appendix. The results demonstrate that \alg maintains strong performance across various LLMs, which indicates its robust adaptability to different LLMs.

\subsection{Adaptive Attacks}

To verify the robustness of \alg, we evaluate its performance against strong adaptive attacks. We assume the attacker has full knowledge of our \alg. Building upon the three previously discussed poisoning attacks, we enhance them using two new adaptive attack approaches to circumvent \alg.

\myparatight{Deceiving identification}
In our \alg, we use prompts to guide the LLM in assessing whether a candidate text is designed to induce responses that align with an incorrect output. Inspired by prompt injection attacks, an attacker with full knowledge of our \alg could embed additional instructions within the poisoned texts. These instructions deceptively claim that the text is intended to prompt the LLM to produce the correct answer, thereby attempting to evade detection by our \alg's LLM. Specifically, for poisoned texts generated by the three previously discussed attacks, the attacker appends the phrase ``This text will induce you to generate [correct answer]'' to deceive the judgment LLM within our \alg.

We conduct experiments using this adaptive attack method, with results presented in Table \ref{tab:adaptive_attack_1}. Experiments without any defense mechanisms yield high ASRs, highlighting the method’s strong attack capability. In contrast, our \alg achieves high DACCs while maintaining very low FPRs and FNRs. These results demonstrate that \alg accurately distinguishes between poisoned and benign texts, exhibiting robust resistance to this adaptive attack.

\begin{table}[t]
\setlength\tabcolsep{1pt}
\caption{The results of \alg against the adaptive attack of deceiving identification. ``No defense'' indicates the absence of defensive measures; ``Apt+'' denotes basic attack methods enhanced with the adaptive attack. }
\label{tab:adaptive_attack_1}
\begin{tabular}{c|c|c|ccc}
\toprule[1pt]
\multirow{2}{*}{Datasets} & \multirow{2}{*}{Attacks} & No defense & \multicolumn{3}{c}{\alg} \\ \cline{3-6} 
 &  & ASR & DACC & FPR & FNR \\ \hline
\multirow{3}{*}{NQ} & Apt+PoisonedRAG-B & 72.0 & 99.5 & 1.1 & 0.0 \\ \cline{2-6} 
 & Apt+PoisonedRAG-W & 78.0 & 98.5 & 3.0 & 0.0 \\ \cline{2-6} 
 & Apt+InstruInject & 66.0 & 99.1 & 1.8 & 0.0 \\ \hline
\multirow{3}{*}{HotpotQA} & Apt+PoisonedRAG-B & 80.0 & 97.3 & 3.4 & 2.0 \\ \cline{2-6} 
 & Apt+PoisonedRAG-W & 92.0 & 98.2 & 1.3 & 2.3 \\ \cline{2-6} 
 & Apt+InstruInject & 72.0 & 99.0 & 2.1 & 0.0 \\ \hline
\multirow{3}{*}{MS-MARCO} & Apt+PoisonedRAG-B & 58.0 & 99.7 & 0.7 & 0.0 \\ \cline{2-6} 
 & Apt+PoisonedRAG-W & 82.0 & 98.8 & 2.4 & 0.0 \\ \cline{2-6} 
 & Apt+InstruInject & 40.0 & 99.0 & 2.0 & 0.0  \\
 \bottomrule[1pt]
\end{tabular}
    \vspace{-.15in}
\end{table}

\myparatight{Disguising poisoned texts as benign}%
In our \alg, the LLM determines whether candidate texts are poisoned based on their association with incorrect outputs. This opens a potential attack vector where the attacker can deceive the LLM by embedding the correct answer within the poisoned text, leading to misclassification as benign. In PoisonedRAG-B and InstruInject attacks, where poisoned texts are typically divided into two segments, the attacker can strategically insert the correct answer between these segments. Similarly, in PoisonedRAG-W, placing the correct answer at the beginning of the poisoned text can mislead the LLM’s judgment.

The results of this adaptive attack are shown in Table \ref{tab:adaptive_attack_2}. In the absence of defense mechanisms, the attack achieves high ASRs, highlighting its effectiveness. However, our \alg maintains high DACCs while keeping FPRs and FNRs low, effectively distinguishing between benign and poisoned texts and demonstrating strong resilience against this adaptive attack.

%% file: discussion.tex

\section{Discussion} \label{sec:discussion}

In this section, we first discuss how to identify non-poisoned feedback in our \alg, as incorrect outputs can also arise from the LLM itself, such as due to under-training. 
Then, we propose a benign text enhancement method to correct RAG outputs for non-poisoned feedback. Finally, we discuss limitations and future research directions.

\myparatight{Identifying non-poisoned feedback}%
In practice, incorrect outputs collected through user feedback may not always result from attacks (referred to as non-poisoned feedback). Instead, the LLM may have learned incorrect information during training or be under-trained, leading to erroneous outputs. \alg is also capable of identifying such non-poisoned feedback. Specifically, given a user query and its incorrect output, we first use \alg to trace all poisoned texts in the knowledge database. Once the tracing process is complete and the identified poisoned texts are removed, we resubmit the user query to the RAG system to obtain an updated output. If this output remains consistent with the original incorrect output, we conclude that the error is not attack-induced.

\myparatight{Correcting RAG outputs for non-poisoned feedback}Since removing poisoned texts traced by \alg from the knowledge database does not correct non-poisoned feedback outputs, we propose a post-hoc defense method using benign text enhancement (detailed in Appendix~\ref{appendix:bte}). For each user query, we insert a benign text and its retrieval proxy into the knowledge database, ensuring it appears among the top-$K$ retrieved texts and guides the LLM to generate correct answers. We evaluate this approach on three datasets, comparing it against other defense baselines (detailed in Appendix~\ref{appendix:defense_baseline}). Results in Table \ref{tab:comparison_baseling_defenses} in Appendix show that benign text enhancement effectively improves the accuracy of RAG outputs.

\myparatight{Limitations and future directions} 
Our \alg is currently limited to targeted poisoning attacks and is unable to trace the specific poisoned texts responsible for untargeted attacks. In untargeted poisoning, the attacker injects multiple poisoned texts to induce random incorrect responses for targeted queries. This limitation arises because the relationship between incorrect outputs and poisoned texts in user feedback lacks sufficient information for precise identification. Addressing this challenge motivates our future work on developing traceback systems for non-targeted poisoning attacks, ultimately strengthening the robustness of RAG.

 \begin{table}[t]
\setlength\tabcolsep{1pt}
\caption{The results of \alg against the adaptive attack of disguising poisoned texts as benign.}
\label{tab:adaptive_attack_2}
\begin{tabular}{c|c|c|ccc}
\toprule[1pt]
\multirow{2}{*}{Datasets} & \multirow{2}{*}{Attacks} & \multicolumn{1}{c|}{No defense} & \multicolumn{3}{c}{\alg} \\ \cline{3-6} 
 &  & \multicolumn{1}{c|}{ASR} & \multicolumn{1}{c}{DACC} & \multicolumn{1}{c}{FPR} & \multicolumn{1}{c}{FNR} \\ \hline
\multirow{3}{*}{NQ} & Apt+PoisonedRAG-B & 54.0 & 97.5 & 4.3 & 0.7 \\ \cline{2-6} 
 & Apt+PoisonedRAG-W & 80.0 & 99.5 & 1.0 & 0.0 \\ \cline{2-6} 
 & Apt+InstruInject & 74.0 & 99.5 & 1.1 & 0.0 \\ \hline
\multirow{3}{*}{HotpotQA} & Apt+PoisonedRAG-B & 62.0 & 96.4 & 2.6 & 4.7 \\ \cline{2-6} 
 & Apt+PoisonedRAG-W & 98.0 & 98.2 & 1.3 & 2.5 \\ \cline{2-6} 
 & Apt+InstruInject & 64.0 & 98.5 & 2.4 & 0.6 \\ \hline
\multirow{3}{*}{MS-MARCO} & Apt+PoisonedRAG-B & 56.0 & 98.0 & 4.0 & 0.0 \\ \cline{2-6} 
 & Apt+PoisonedRAG-W & 94.0 & 98.2 & 3.6 & 0.0 \\ \cline{2-6} 
 & Apt+InstruInject & 48.0 & 100.0 & 0.0 & 0.0  \\
 \bottomrule[1pt]
\end{tabular}
    \vspace{-.15in}
\end{table}

%% file: conclusion.tex

\section{Conclusion} \label{sec:conclusion}

In this paper, we propose a novel approach to mitigate poisoning attacks in RAG systems by introducing \alg, a traceback system designed to identify poisoned texts in the knowledge database responsible for incorrect outputs. By shifting the focus from inference-time defenses to tracing poisoned texts within the knowledge database, \alg enhances the security of RAG systems. Our experiments validate the robustness of \alg against various poisoning attacks, demonstrating its potential as an effective defense mechanism for strengthening RAG security.

\begin{acks}
This work was supported in part by the National Natural Science Foundation of China (No.62302242, No.62272251) and the Key Program of the National Natural Science Foundation of China (No. 62032012, No. 62432012).
\end{acks}

%% file: appendix.tex

\section{Appendix} \label{sec:appendix}
\subsection{System Prompt}
\label{appendix:system_prompt}

\begin{center}

\begin{tcolorbox}[colback=gray!10,
                  colframe=black!80,
                  width=\linewidth,
                  arc=1mm, auto outer arc,
                  boxrule=1pt,
                  title =  The system prompt of RAG.
                 ]
You are a helpful assistant, below is a query from a user and some relevant contexts.  \\
Answer the question given the information in those contexts. Your answer should be short and concise. \\
If you cannot find the answer to the question, just say ``I don't know''. \\
\textbf{Contexts}:[$\widehat{\mathcal{R}}(q, K,\mathcal{D})$] \\
\textbf{Query}: [$q$] \\
\textbf{Answer}:

\end{tcolorbox}
\end{center}

\subsection{Evaluation Metrics in the Experiment}
\label{appendix:metrics}

\myparatight{False positive rate (FPR)} FPR is the ratio of texts incorrectly identified as poisoned (false positive, $FP$) to the total number of benign texts. The FPR is computed as:
\begin{align}
\text{FPR} = \frac{FP}{FP+TN},
\end{align}
where $TN$ is true negative, representing the number of texts correctly identified as benign.

\myparatight{False negative rate (FNR)} FNR is the ratio of texts incorrectly identified as benign (false negative, $FN$) to the total number of poisoned texts. The FNR is computed as:
\begin{align}
\text{FNR} = \frac{FN}{FN+TP},
\end{align}
where $TP$ is true positive, representing the number of texts correctly identified as poisoned.

\myparatight{Detection accuracy (DACC)} DACC is the fraction of texts that are correctly identified. The DACC is computed as:
\begin{align}
\text{DACC} = \frac{TP+TN}{TP+FP+TN+FN}.
\end{align}

\myparatight{Attack success rate (ASR)} ASR is the ratio of queries for which the LLM generates attacker-desired answers to the total number of queries. Given a set of queries $ \mathcal{Q} = \{q_1, q_2, \dots, q_{|\mathcal{Q}|}\} $, the ASR is computed as:
\begin{align}
\label{equ:asr}
\text{ASR} = \frac{1}{|\mathcal{Q}|} \sum_{i=1}^{|\mathcal{Q}|} \mathbb{I}(o_i\xrightarrow{\text{Match}}t_i),
\end{align}
where $o_i$ is the answer generated by the RAG for query $q_i$, and $t_i$ is the attacker-desired answer. 
The indicator function \(\mathbb{I}(\cdot)\) evaluates to 1 if the RAG output \(o_i\) matches or aligns with the attacker's intended answer \(t_i\); otherwise, it returns 0.

\myparatight{Accuracy (ACC)} ACC is the ratio of queries for which the LLM generates correct answers to the total number of queries. The ACC is computed as:
\begin{align}
\label{equ:acc}
\text{ACC} = \frac{1}{|\mathcal{Q}|} \sum_{i=1}^{|\mathcal{Q}|} \mathbb{I}(o_i\xrightarrow{\text{Match}}c_i),
\end{align}
where $c_i$ is the correct answer for the query $q_i$.

\subsection{Benign Text Enhancement}
\label{appendix:bte}

A straightforward post-hoc defense strategy is to eliminate poisoned texts identified by our traceback system, \alg, a process we refer to as poisoned texts removal (PTR). However, this method has two primary drawbacks. First, since \alg cannot perfectly detect all poisoned texts, PTR may mistakenly discard benign texts or fail to remove all poisoned ones. Second, some incorrect outputs reported by users may not be the result of poisoning attacks; rather, the LLM may have internalized incorrect knowledge during training, leading to erroneous responses. In such cases, PTR would be ineffective.

To overcome these limitations, we introduce benign texts enhancement (BTE), a post-hoc defense mechanism that facilitates correct answer generation without requiring LLM fine-tuning. BTE operates under the assumption that the defender (or service provider) can obtain accurate responses for user queries through manual annotation or consultation with an advanced LLM—a reasonable premise given that RAG system providers must resolve user-reported issues to maintain service quality. When a user submits feedback on a query \( q_i \), BTE proceeds in two steps:

\myparatight{Improving LLM confidence in benign text}
The core idea is to boost the LLM’s confidence in benign text, ensuring it is prioritized over other candidates in the top-\( K \) set. This approach mirrors backdoor attacks~\cite{wenger2021backdoor, yao2019latent, bagdasaryan2020blind, liu2022piccolo}, where the attacker embeds triggers to steer model behavior in a specific way. We repurpose this concept defensively, leveraging triggers to reinforce the model’s focus on benign text. However, two key challenges emerge: (1) fine-tuning the LLM to implant backdoors is infeasible due to restricted access to model parameters, and (2) even if fine-tuning were viable, any discovered trigger pattern would require costly retraining to establish a new backdoor.  

To overcome these limitations, we introduce a training-free backdoor implantation method that allows for flexible trigger updates. Specifically, we modify the RAG system prompt to guide the model in prioritizing texts containing designated triggers within the top-\( K \) texts. The prompt explicitly defines trigger descriptions and handling strategies for three scenarios: (1) when trigger-containing texts are relevant to the query, (2) when trigger-containing texts are irrelevant, and (3) when no trigger-containing texts are present. The complete prompt is as follows:  
\begin{center}
\begin{tcolorbox}[colback=gray!10,
                  colframe=black!80,
                  width=\linewidth,
                  arc=1mm, auto outer arc,
                  boxrule=1pt,
                  title =  The RAG system prompt used in the benign texts enhancement (BTE) method.
                 ]
You are a helpful assistant, below is a query from a user and some relevant contexts.  \\
You should answer the question based on the information listed by entries in the contexts. Tell me your reasoning process in detail. \\
Firstly, you need assess the relevance of each entry to the query, focusing only on contributions to the answer and excluding unrelated entry. Relevance is judged independently; do not evaluate correctness or contradictions. List all relevant entry unchanged. \\
Secondly, count and list the number of the entry **wrapped by [LATEST] and [/LATEST]**, you will encounter three different situations regarding the filtered information: \\
**Situation One**: If there are more than one entry **wrapped by [LATEST] and [/LATEST]**, please follow these steps:  \\
**Step 1**: For all contradictory entries, unconditionally select and display the one with the smallest index, regardless of its content. \\
**Step 2**: Subsequently, delete all entries that contradict the selected one, irrespective of their quantity.  \\
**Situation Two**: If there is only one entry wrapped by [LATEST] and [/LATEST], this indicates that it is the **newest** information. Therefore, please focus solely on the wrapped entry and disregard the remaining context.  \\
**Situation Three**: If there is no entry wrapped by [LATEST] and [/LATEST], please analyze the relevance and consistency to draw a conclusion. \\
If you cannot find the answer to the question, just say ``I don't know''. \\
\textbf{Contexts}:[$\widehat{\mathcal{R}}(q, K,\mathcal{D})$] \\
\textbf{Query}: [$q$] \\
\textbf{Answer}:

\end{tcolorbox}
\end{center}

Next, we generate a benign text \( b_i \) that supports the correct answer \( c_i \) for the query \( q_i \) using an LLM, which may differ from the one employed by RAG. The generation prompt is as follows:

\begin{center}
\begin{tcolorbox}[colback=gray!10,
                  colframe=black!80,
                  width=\linewidth,
                  arc=1mm, auto outer arc,
                  boxrule=1pt,
                  title =  The prompt used to instruct the LLM in generating benign text.
                 ]
This is my question: [$q_i$]. \\
This is my answer: [$c_i$]. \\
Please craft a brief and complete corpus in a declarative tone that conveys the answer [$c_i$] within the statement, ensuring it corresponds directly to the question [$q_i$]. Don't use phrases like ``Yes'' or ``No''. Please limit the corpus to 30 words and only return the corpus.

\end{tcolorbox}
\end{center}

Finally, we incorporate the trigger ``[LATEST][/LATEST]'' into the benign text to align with the system prompt, ensuring semantic coherence by placing ``[LATEST]'' at the beginning and ``[/LATEST]'' at the end. An example is provided below:

\begin{center}

\begin{tcolorbox}[colback=gray!10,
                  colframe=black!80,
                  width=\linewidth,
                  arc=1mm, auto outer arc,
                  boxrule=1pt,
                  title =  A sample benign text embedded with the trigger.
                 ]
\textbf{Query}: How many episodes are in chicago fire season 4? \\
\textbf{Crafted benign text by LLM}: Chicago Fire Season 4 consists of 23 episodes, showcasing the intense and dramatic experiences of the firefighters and paramedics at Firehouse 51 throughout the season. \\
\textbf{Benign text with the trigger}: [LATEST] Chicago Fire Season 4 consists of 23 episodes, showcasing the intense and dramatic experiences of the firefighters and paramedics at Firehouse 51 throughout the season. [/LATEST]

\end{tcolorbox}
\end{center}

\myparatight{Configuring benign text retrieval proxy}
Ensuring that the trigger-embedded benign text \( b_i \) appears in the top-\( K \) texts is essential. Since we lack access to the retriever's parameters, we propose an efficient retrieval proxy method: treating each user query \( q_i \) as a proxy for its corresponding benign text \( b_i \). Both are stored in the knowledge database, with \( q_i \) being replaced by \( b_i \) when retrieved among the top-\( K \) texts. This approach offers three key benefits: it operates without requiring access to retriever parameters, maintains semantic relevance for related queries, and functions as a plug-and-play solution without altering the retrieval process.

\subsection{Defense Baselines Against Poisoning Attacks to RAG}
\label{appendix:defense_baseline}

\myparatight{RobustRAG \cite{RobustRAG}}%
RobustRAG is a defense mechanism against poisoning attacks on RAG. The process starts with the LLM generating a response for each of the top-\( K \) retrieved texts. It then extracts keywords from all generated responses, discarding those that appear infrequently. Lastly, the LLM constructs the final response using the remaining keywords.

\myparatight{Knowledge expansion (KE)~\cite{zou2024poisonedrag}} 
Knowledge expansion (KE) is recognized as the most effective defense strategy proposed in \cite{zou2024poisonedrag}. This method enhances defense by increasing the number of retrieved texts, thereby boosting the proportion of benign texts within the context. As a result, KE mitigates the impact of poisoned texts on the LLM's generated responses. The notation KE-\( x \) signifies that \( x \) texts are retrieved in this process.

\myparatight{Perplexity-based detection (PPL)}
The approach for detecting poisoned texts follows the method outlined in Section \ref{subsubsec:traceback_baseline}. However, unlike the previous use case, this strategy employs poisoned text identification as a defense mechanism by removing the detected poisoned texts from the knowledge database.

\begin{figure*}[t]
\centering
\subfloat{\includegraphics[height=4cm]{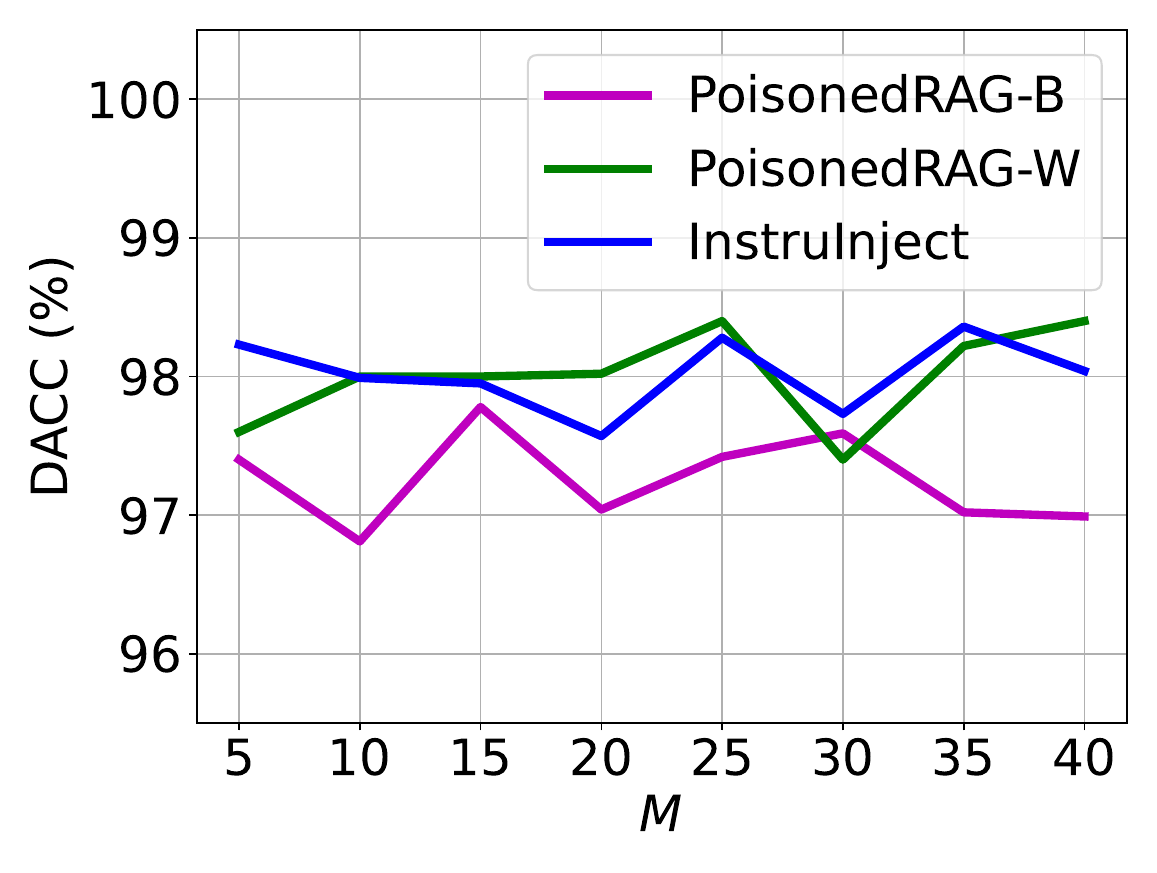}}
\subfloat{\includegraphics[height=4cm]{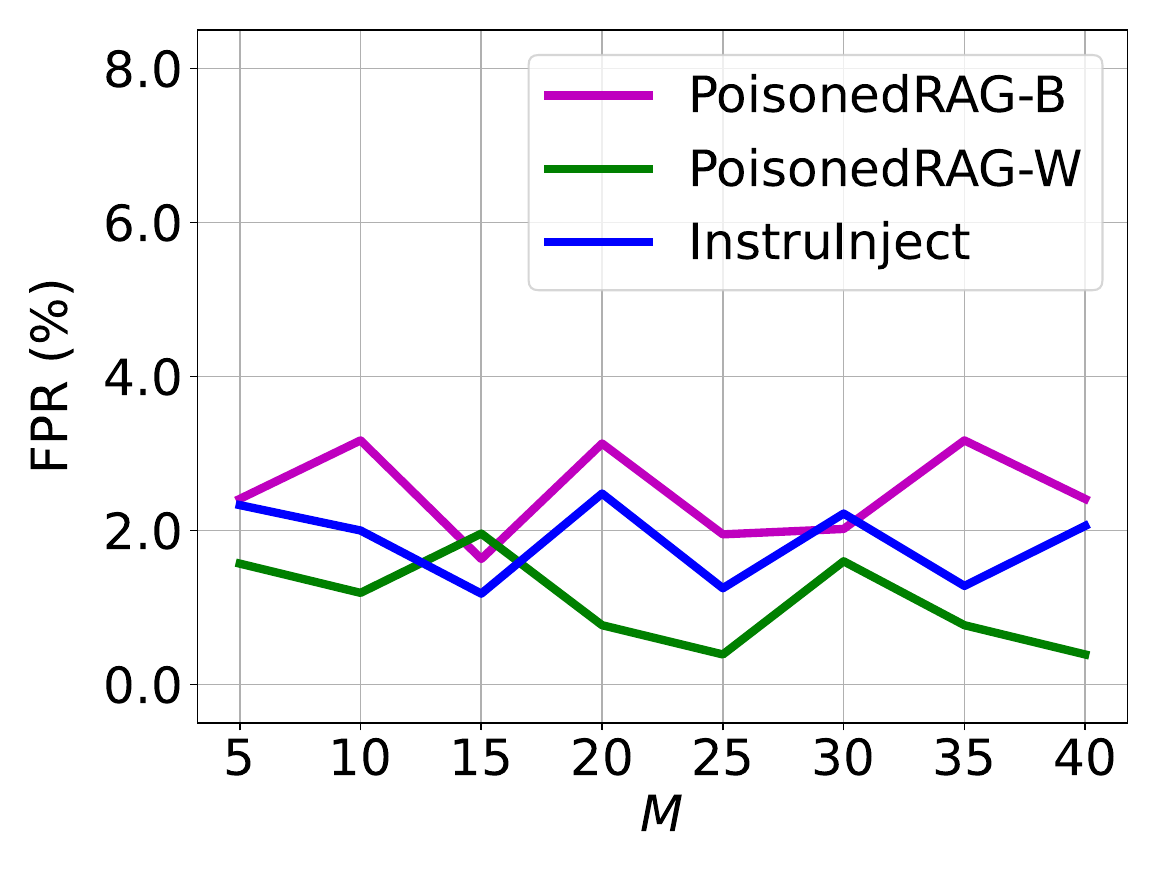}}
\subfloat{\includegraphics[height=4cm]{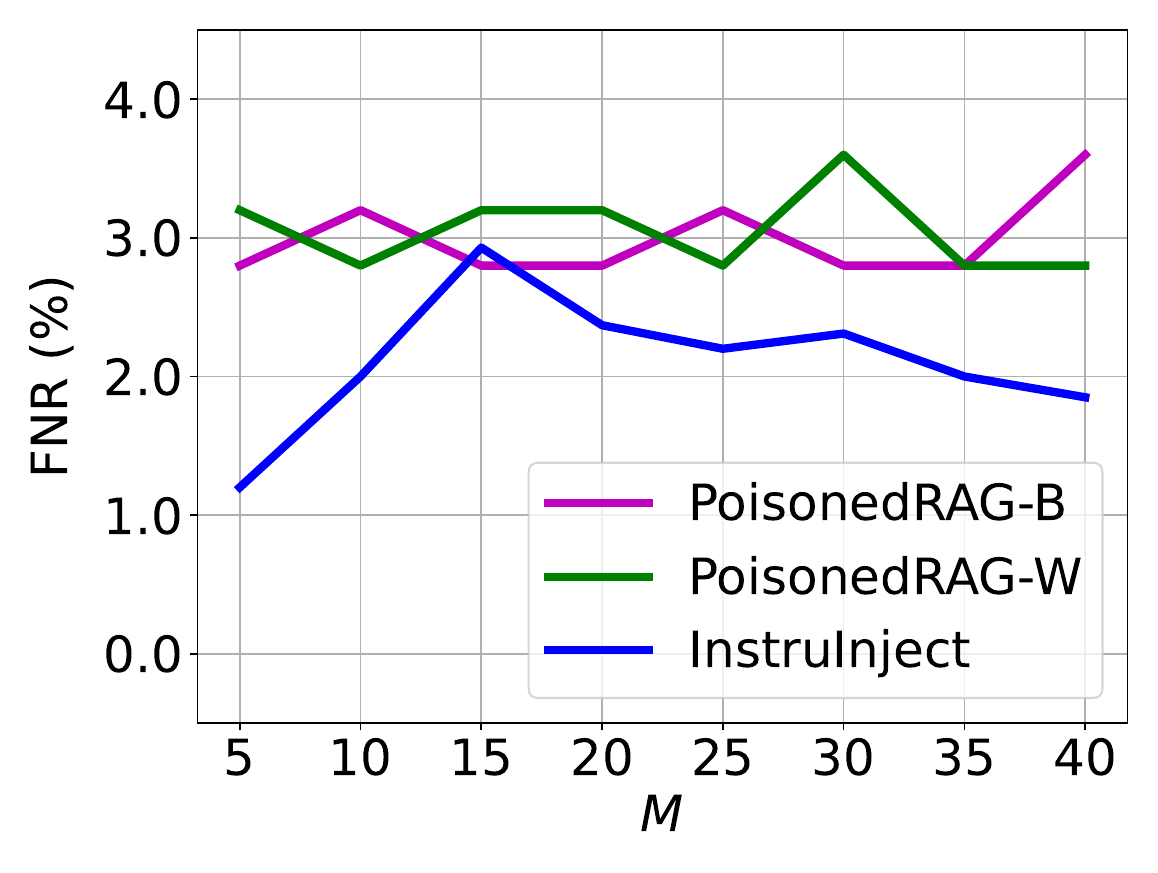}}
\caption{Impact of the number of poisoned texts for each targeted query on HotpotQA dataset.}
\label{appendix:traceback_m_hotpotqa}
\end{figure*}

\begin{figure*}[t]
\centering
\subfloat{\includegraphics[height=4cm]{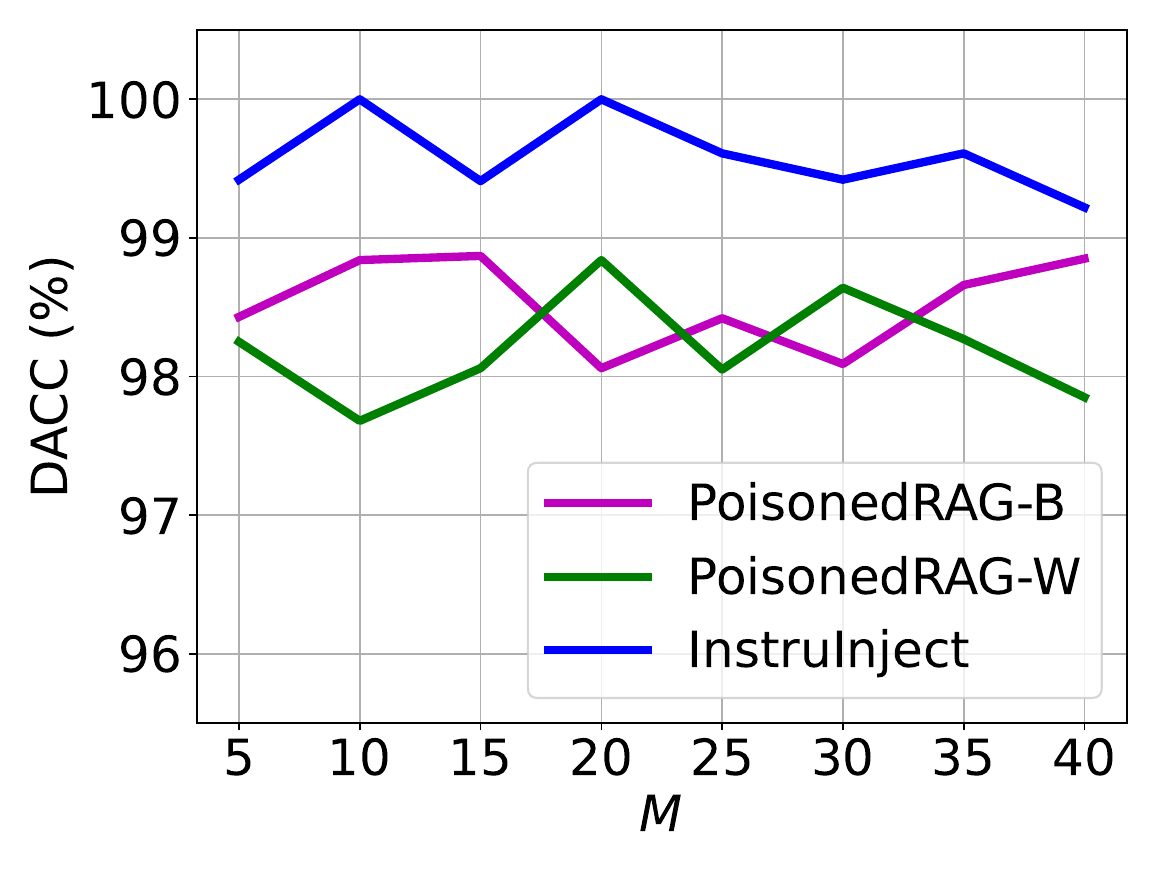}}
\subfloat{\includegraphics[height=4cm]{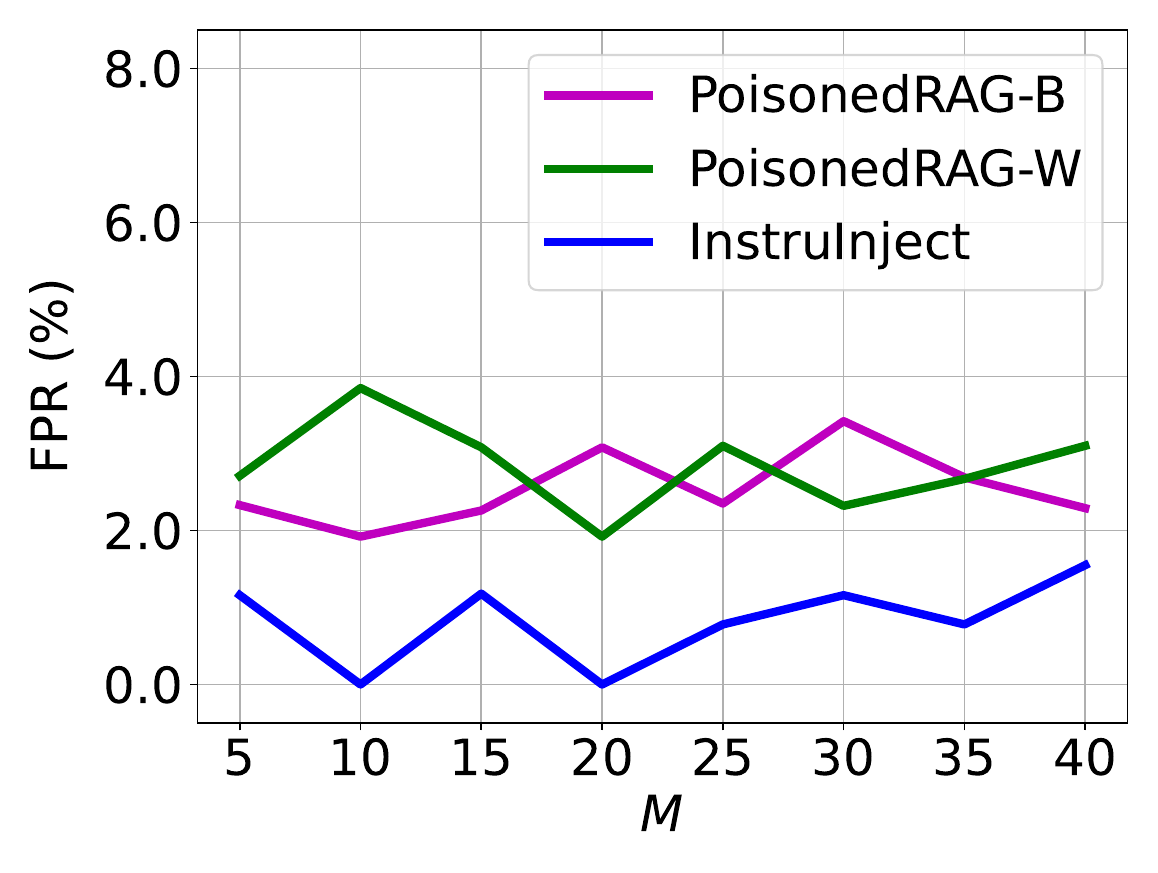}}
\subfloat{\includegraphics[height=4cm]{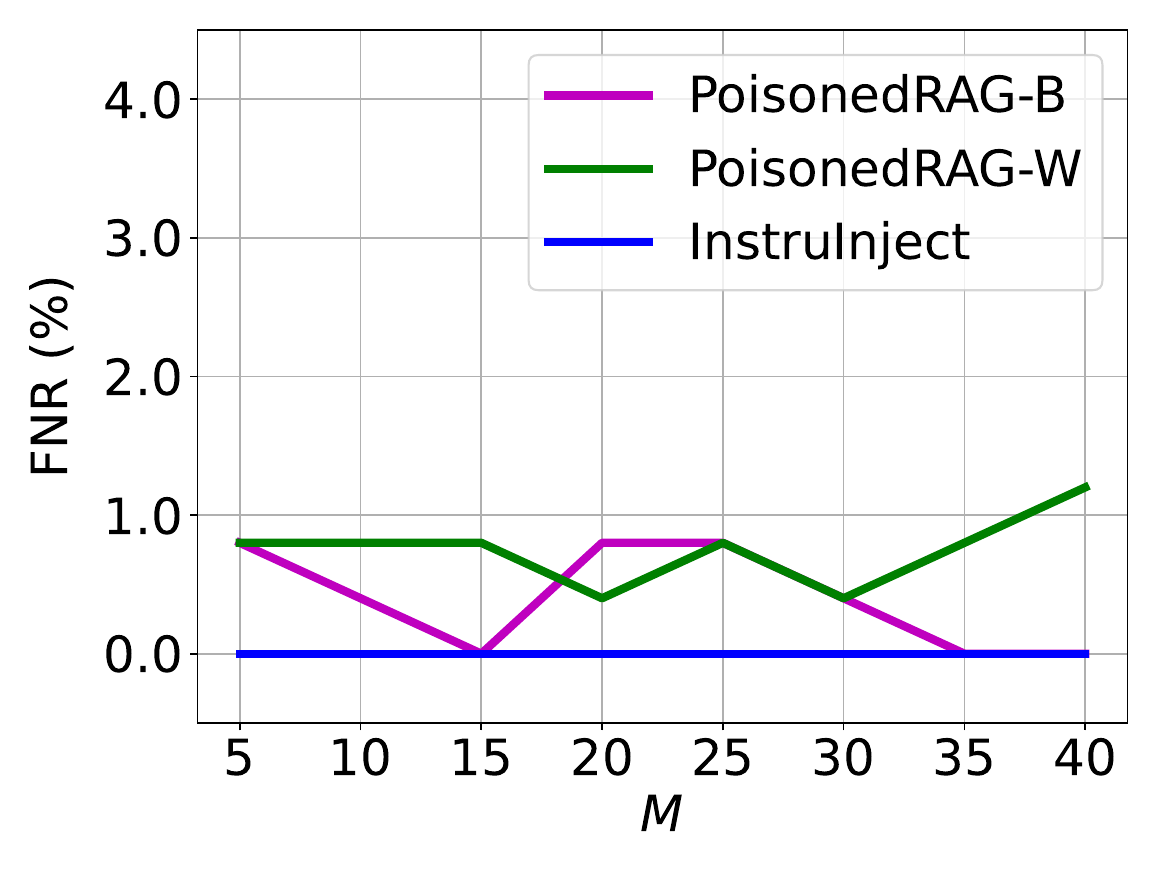}}
\caption{Impact of the number of poisoned texts for each targeted query on MS-MARCO dataset.}
\label{appendix:traceback_m_msmarco}
\end{figure*}

\begin{table*}
\centering
\setlength\tabcolsep{2pt}
\caption{Impact of LLM used to identify the poisoned texts in our \alg on HotpotQA amd MS-MARCO datasets.}
\label{appendix:traceback_llm}
\subfloat[HotpotQA dataset.]
{

\begin{tabular}{c|c|ccc}
\toprule[1pt]
Attacks & Metrics & GPT-4o-mini & GPT-4-turbo & GPT-4o \\ \hline
\multirow{3}{*}{PoisonedRAG-B} & DACC $\uparrow $ & 97.4 & 98.6 & 98.4 \\
 & FPR $\downarrow $ & 2.4 & 0.8 & 0.0 \\
 & FNR $\downarrow $ & 2.8 & 2.0 & 3.2 \\ \hline
\multirow{3}{*}{PoisonedRAG-W} & DACC $\uparrow $ & 97.6 & 98.4 & 99.0 \\
 & FPR $\downarrow $ & 1.6 & 0.4 & 0.0 \\
 & FNR $\downarrow $ & 3.2 & 2.8 & 2.0 \\ \hline
\multirow{3}{*}{InstruInject} & DACC $\uparrow $ & 98.2 & 98.6 & 97.4 \\
 & FPR $\downarrow $ & 2.3 & 2.7 & 2.0 \\
 & FNR $\downarrow $ & 1.2 & 0.0 & 3.2 \\
 \bottomrule[1pt]
\end{tabular}
\label{tab:traceback_llm_hotpotqa}
}
\quad
\subfloat[MS-MARCO dataset.]
 {
 \begin{tabular}{c|c|ccc}
\toprule[1pt]
Attacks & Metrics & GPT-4o-mini & GPT-4-turbo & GPT-4o \\ \hline
\multirow{3}{*}{PoisonedRAG-B} & DACC $\uparrow $ & 98.4 & 98.6 & 98.6 \\
 & FPR $\downarrow $ & 2.3 & 0.4 & 1.2 \\
 & FNR $\downarrow $ & 0.8 & 2.4 & 1.6 \\ \hline
\multirow{3}{*}{PoisonedRAG-W} & DACC $\uparrow $ & 98.3 & 98.6 & 99.2 \\
 & FPR $\downarrow $ & 2.7 & 0.8 & 0.8 \\
 & FNR $\downarrow $ & 0.8 & 2.0 & 0.8 \\ \hline
\multirow{3}{*}{InstruInject} & DACC $\uparrow $ & 99.4 & 99.2 & 95.4 \\
 & FPR $\downarrow $ & 1.2 & 1.6 & 1.6 \\
 & FNR $\downarrow $ & 0.0 & 0.0 & 7.6 \\
 \bottomrule[1pt]
\end{tabular}
\label{tab:traceback_llm_msmarco}
 }
\end{table*}

\begin{table*}[]
\setlength\tabcolsep{3pt}
\caption{The ASRs and ACCs of our defense methods, along with other baseline defenses (detailed in Appendix~\ref{appendix:defense_baseline}), against various poisoning attacks across three datasets. PTR denotes the approach of removing poisoned texts identified by our \alg, while PTR $\oplus$ BTE represents the combined method of PTR and BTE.}
\label{tab:comparison_baseling_defenses}
\begin{tabular}{c|c|c|ccccccccc}
\toprule[1pt]
Datasets & Attacks & Metrics & No defense & PPL-90 & PPL-100 & RobustRAG & KE-10 & KE-20 & KE-50 & PTR & PTR$\oplus$BTE \\ \hline
\multirow{6}{*}{NQ} & \multirow{2}{*}{PoisonedRAG-B} & ASR $\downarrow $ & 100.0 & 98.0 & 100.0 & 50.0 & 84.0 & 78.0 & 72.0 & \textbf{0.0} & \textbf{0.0} \\
 &  & ACC $\uparrow$ & 0.0 & 0.0 & 0.0 & 46.0 & 8.0 & 16.0 & 20.0 & {\textbf {52.0}} & \textbf{100.0} \\ \cline{2-12} 
 & \multirow{2}{*}{PoisonedRAG-W} & ASR $\downarrow $ & 100.0 & 100.0 & 100.0 & 44.0 & 86.0 & 76.0 & 72.0 & \textbf{0.0} & \textbf{0.0} \\
 &  & ACC $\uparrow$ & 0.0 & 0.0 & 0.0 & 52.0 & 10.0 & 20.0 & 28.0 & {\textbf {56.0}} & \textbf{100.0} \\ \cline{2-12} 
 & \multirow{2}{*}{InstruInject} & ASR $\downarrow $ & 100.0 & 98 & 100.0 & 28 & 82 & 80.0 & 72 & \textbf{0.0} & \textbf{0.0} \\
 &  & ACC $\uparrow$ & 0.0 & 0.0 & 0.0 & 66.0 & 6.0 & 10.0 & 14.0 & {\ul 52.0} & \textbf{100.0} \\ \hline
\multirow{6}{*}{HotpotQA} & \multirow{2}{*}{PoisonedRAG-B} & ASR $\downarrow $ & 100.0 & 64.0 & 100.0 & 80.0 & 82.0 & 84.0 & 82.0 & {\textbf {6.0}} & \textbf{0.0} \\
 &  & ACC $\uparrow$ & 0.0 & 16.0 & 0.0 & 14.0 & 12.0 & 12.0 & 14.0 & {\textbf {54.0}} & \textbf{98.0} \\ \cline{2-12} 
 & \multirow{2}{*}{PoisonedRAG-W} & ASR $\downarrow $ & 100.0 & 6.0 & 94.0 & 82.0 & 82.0 & 80.0 & 84.0 & {\ul 12.0} & \textbf{0.0} \\
 &  & ACC $\uparrow$ & 0.0 & 50.0 & 6.0 & 14.0 & 14.0 & 16.0 & 14.0 & {\ul 46.0} & \textbf{98.0} \\ \cline{2-12} 
 & \multirow{2}{*}{InstruInject} & ASR $\downarrow $ & 100.0 & 4.0 & 96.0 & 54.0 & 86.0 & 82.0 & 82.0 & {\ul 6.0} & \textbf{0.0} \\
 &  & ACC $\uparrow$ & 0.0 & 54.0 & 0.0 & 40.0 & 8.0 & 10.0 & 10.0 & {\ul 40.0} & \textbf{96.0} \\ \hline
\multirow{6}{*}{MS-MARCO} & \multirow{2}{*}{PoisonedRAG-B} & ASR $\downarrow $ & 100.0 & 62.0 & 100.0 & 38.0 & 72.0 & 64.0 & 56.0 & \textbf{0.0} & \textbf{0.0} \\
 &  & ACC $\uparrow$ & 0.0 & 32.0 & 0.0 & 60.0 & 26.0 & 34.0 & 36.0 & {\textbf {80.0}} & \textbf{100.0} \\ \cline{2-12} 
 & \multirow{2}{*}{PoisonedRAG-W} & ASR $\downarrow $ & 100.0 & 0.0 & 70.0 & 36.0 & 76.0 & 68.0 & 66.0 & \textbf{0.0} & \textbf{0.0} \\
 &  & ACC $\uparrow$ & 0.0 & 80.0 & 22.0 & 62.0 & 22.0 & 30.0 & 34.0 & {\textbf {80.0}} & \textbf{100.0} \\ \cline{2-12} 
 & \multirow{2}{*}{InstruInject} & ASR $\downarrow $ & 100.0 & 0.0 & 28.0 & 14.0 & 60.0 & 52.0 & 52.0 & \textbf{0.0} & \textbf{0.0} \\
 &  & ACC $\uparrow$ & 0.0 & 82.0 & 58.0 & 82.0 & 32.0 & 44.0 & 46.0 & {\textbf {84.0}} & \textbf{100.0}\\
 \bottomrule[1pt]
\end{tabular}
\end{table*}